%% file: eben.tex
\documentclass[conference]{IEEEtran}
\IEEEoverridecommandlockouts
\usepackage{cite}
\usepackage{amsmath,amssymb,amsfonts}
\usepackage{algorithmic}
\usepackage{graphicx}
\usepackage{textcomp}
\usepackage{xcolor}
\usepackage{hyperref}
\usepackage{diagbox}
\usepackage{subcaption}
\usepackage{adjustbox}

\hypersetup{pdfauthor={Hauret, Julien and Joubaud, Thomas and Zimpfer, Véronique and Bavu, Eric},
            pdftitle={Configurable EBEN},
            pdfsubject={Extreme Bandwidth Extension Network to enhance body-conducted speech capture},
            pdfkeywords={Speech enhancement} {PQMF-banks} {Bandwidth extension} {Frugal AI} {Signal Processing} {Body-Conduction Microphones},}

\let\orgtilde\tilde
\def\tilde#1{\orgtilde{\kern0pt #1}}

\def\BibTeX{{\rm B\kern-.05em{\sc i\kern-.025em b}\kern-.08em
    T\kern-.1667em\lower.7ex\hbox{E}\kern-.125emX}}
\begin{document}

\title{Configurable EBEN: Extreme Bandwidth Extension Network to enhance body-conducted speech capture}

\author{
\IEEEauthorblockN{1\textsuperscript{st} Hauret Julien}
\IEEEauthorblockA{\textit{Laboratoire de Mécanique des Structures et des Systèmes Couplés} \\
\textit{Conservatoire national des arts et métiers, HESAM Université }\\
Paris, France\\
ORCID : 0000-0002-1512-2487\\
julien.hauret@lecnam.net}
\and
\IEEEauthorblockN{2\textsuperscript{nd} Joubaud Thomas}
\IEEEauthorblockA{\textit{Department of Acoustics and Soldier Protection} \\
\textit{French-German Research Institute of Saint-Louis (ISL)}\\
Saint-Louis, France \\
ORCID : 0000-0002-5324-8785\\
thomas.joubaud@isl.eu}
\and
\IEEEauthorblockN{3\textsuperscript{rd} Zimpfer Véronique}
\IEEEauthorblockA{\textit{Department of Acoustics and Soldier Protection} \\
\textit{French-German Research Institute of Saint-Louis (ISL)}\\
Saint-Louis, France  \\
ORCID : 0000-0002-7852-1928\\
veronique.zimpfer@isl.eu}
\and
\IEEEauthorblockN{4\textsuperscript{th} Bavu Éric}
\IEEEauthorblockA{\textit{Laboratoire de Mécanique des Structures et des Systèmes Couplés} \\
\textit{Conservatoire national des arts et métiers, HESAM Université }\\
Paris, France \\
ORCID : 0000-0001-6395-634X\\
eric.bavu@lecnam.net}
}

\maketitle

\begin{abstract}
This paper presents a configurable version of Extreme Bandwidth Extension Network (EBEN), a Generative Adversarial Network (GAN) designed to improve audio captured with body-conduction microphones. We show that although these microphones significantly reduce environmental noise, this insensitivity to ambient noise happens at the expense of the bandwidth of the speech signal acquired by the wearer of the devices. The obtained captured signals therefore require the use of signal enhancement techniques to recover the full-bandwidth speech. EBEN leverages a configurable multiband decomposition of the raw captured signal. This decomposition allows the data time domain dimensions to be reduced and the full band signal to be better controlled. The multiband representation of the captured signal is processed through a  U-Net-like model, which combines feature and adversarial losses to generate an enhanced speech signal. We also benefit from this original representation in the proposed configurable discriminators architecture. The configurable EBEN approach can achieve state-of-the-art enhancement results on synthetic data with a lightweight generator that allows real-time processing.
\end{abstract}


\section{Introduction}
\label{sec:intro}

Capturing speech involves the use of microphones to transform mechanical vibrations into an electrical signal, later digitalized and eventually used for radio communications. Under quiet conditions, using an airborne-sound microphone near the speaker's lips is the most appropriate way to capture clean speech. Nevertheless, in presence of ambient noise generated by sources contaminating the sound scene, the speech signal of interest is altered by the acoustic environment, which also contributes to air molecules vibration. This situation -- which reduces the intelligibility of communications -- is frequently encountered in industry, on the battlefield or in strong winds. In extreme cases, operators are even unable to communicate.

Before using any speech enhancement technique, it is worth pondering the best mechanical signal to rely on in noisy conditions. There are other choices besides recording airborne sound pressure, such as the body-conducted inner vibrations of the speaker. The human body is not as easily moved by environmental noise as ambient air, due to the high damping of the transmitted sound wave in the tissues. Therefore, capturing inner tissues' vibrations caused by the vocal tract near the speaker's head has great potential for improving the signal to noise ratio when recording speech in noisy environments. This can be performed with noise-resilient body-conduction microphones (BCMs), which allow sensing the internal vibrations of the equipped person. This family of unconventional voice pickup systems includes bone conduction transducers \cite{shin2012survey,mcbride2011effect,li2014multisensory,li2022enabling,acker2005speech}, throat microphones \cite{shahina2007mapping,turan2013enhancement} and in-ear microphones mounted in occlusive earplugs offering hearing protection \cite{bos2005speech,bouserhal2017ear,park2019speech,ohlenbusch2022training}. Studies including \cite{mcbride2011effect,bouserhal2017ear} and \cite{casali1996technology} demonstrated that they offer higher quality and intelligibility in noise than conventional capture devices. We also conducted our experiment on Sec.~\ref{sec:innnoisecomparison} to determine when it is preferable to use a BCM over a traditional microphone.

In addition to eliminating external noise pollution, BCMs are less invasive and compatible with helmets, which are often required in noisy environments. Similarly, they are suitable for wearing gas masks or face masks which is not negligible in times of pandemic. In-ear capture devices are also prone to be integrated into hearing protection devices. The protection will isolate the sensor from the external environment, and the wearer's speech capture will be improved. Finally, the broad adoption of true wireless stereo earbuds and bone conduction earphones also benefits the development of inner voice capture. Indeed, those systems are reversible and could be used as BCMs.

Despite many advantages, the usage of BCMs has not yet been democratized. This can be marginally explained by the fact that they are not always necessary (e.g., in a quiet, distant meeting), but mainly because recordings suffer from reduced bandwidth. Indeed, mid and high frequencies are missing due to the intrinsic low-pass characteristics of the biological pathway. Further processing is then necessary to optimize the effective bandwidth of the captured speech. Moreover, other physiological sounds, such as swallowing, blood flow - or any other sound produced by the body - are also picked up by BCMs and represent a new form of noise contaminating speech capture.

In simple terms, speech capture in surrounding noise can be achieved either by using airborne speech with a denoising algorithm or by using a noise-proof body-conduction microphone with bandwidth extension techniques. The latter is a viable solution for critical noise levels ($\geq$ 85dB) when differential microphones or directional boom microphones cannot eliminate high-level surrounding noise. Therefore, this article proposes an extreme bandwidth extension deep neural network for speech signals captured with noise-resilient body-conduction microphones.

Since the desirable system could be a two-way communication device for industrial of military usecases, this entails real-time execution constraints, \emph{i.e.} a short processing time to be indistinguishable from the human ear. Moreover, edge computations are required to guarantee low latency, necessitating a light algorithm. These considerations also match frugal AI requirements. Finally, the developed model should be robust to speaker identity, physiological and residues of external noises that would have infiltrated the microphone.

To meet the expectations of extreme bandwidth extension and related requirements, research like \cite{li2021real,li2022enabling,kong2020hifi} suggests that frugal deep learning is an appropriate approach. Indeed, compared to conventional signal processing methods that can only extend existing frequency content, deep learning can regenerate missing components such as fricatives.  Additionally, deep learning offers the advantage of simultaneous denoising and signal enhancement, eliminating the need for separate denoising procedures. On the other hand, massive deep learning models are not relevant for real-time execution.

Based on the above observations, we developed configurable EBEN, a new deep learning model inspired by a family of lightweight convolutional-based encoder-decoder architecture \cite{defossez2020real,tagliasacchi2020seanet,li2021real,zeghidour2021soundstream,defossez2022high} to infer mid and high frequencies from speech containing only low frequencies (extreme bandwidth extension). We use a generator that maps the degraded speech signal to an enhanced version. This task is called blind speech enhancement because we do not use any external modality (contrary to Seanet \cite{tagliasacchi2020seanet}, which takes advantage of both airborne speech and accelerometer data). EBEN's generator is optimized to produce samples close to the reference while maintaining a certain degree of naturalness at different time scales. We still differ from previous work by using a multiband decomposition performed with Pseudo-Quadrature Mirror Filters (PQMF) \cite{nguyen1994near}. Combined with some hypotheses on addressed degradations, this decomposition is applied to reduce the dimension of input features, which significantly decreases the latency and computational load of the network. In addition, this alternative representation is useful for focusing signal discrimination solely on high frequency bands.

A preliminary version of our research was presented at ICASSP 2023 \cite{hauret2023eben}. The present paper extends the original study by highlighting the usefulness of BCMs in noise, by addressing more diverse and realistic degradations, by discussing the goals and flexibility offered by the configurable aspect of our approach, and by comparing EBEN's latency and memory footprint to other previously published networks. An extensive discussion of the usefulness of common objective metrics is also proposed for the specific task of bandwidth extension, along with a correlation analysis of objective and neural distances with subjective evaluations. The present paper also proposes an extensive discussion of related work. Finally, details of the training strategy, architecture, and statistical analysis of the evaluation survey are presented. The website \url{https://jhauret.github.io/eben} also provides example audio files to listen to and the source code of EBEN.

The body-conduction microphone studied in this paper is an in-ear microphone prototype mounted in an earplug. The few minutes of recordings at our disposal being insufficient to serve for supervised training, we instead analyzed bandwidth loss to simulate in-ear-like degradations on the French Librispeech dataset \cite{pratap2020mls}. Triplet train/dev/test sets of reference and corrupted speech pairs were produced to train our model, and several baselines \cite{kuleshov2017audio,kong2020hifi,tagliasacchi2020seanet,li2021real}. We plan to later release a publicly available dataset of speech capture with BCMs to circumvent the disadvantages of the use of synthetic data, which are discussed in Sec.~\ref{sec:phirandom}.

It is worth noting that focusing on the capture-induced degradation of a specific BCM does not detract from the generality of the proposed approach. This family of sensors consistently degrades speech in a similar manner, acting as a low-pass filter, and adding pysiological and frictional noise that would not be captured by a conventional microphone. Variations occur mainly in cut-off frequency, attenuation, and lack of coherence at certain frequencies. Therefore, a suitable dataset would be sufficient to address any other capture system. It is also noteworthy that preliminary experiments have shown that the EBEN approach performs well at conventional bandwidth extension (\textit{e.g.}~ upsampling 4kHz to 16kHz), which has been confirmed by listening and metrics.

In Sec.~\ref{sec:biblio} we review related previous studies, which also serve as a baseline for our comparisons. In Sec.~\ref{sec:micro}, we show that BCMs are more suitable for recording speech in noise than traditional microphones, present the observed degradation with our in-ear prototype, and describe our protocol for generating synthetic data. Sec.~\ref{sec:eben} provides a brief reminder of the PQMF bank and a detailed presentation of EBEN architecture and loss functions. Sec.~\ref{sec:experiments} describes the training pipeline, experimental results, and compares EBEN to other approaches. The discussion Sec.~\ref{sec:discussion} provides some insights into the EBEN model by discussing why PQMF bands fit well in the context of real-time BCM speech enhancement and the configurable aspects of the architecture. This section also includes a statistical analysis of the correlation between subjective and objective metrics and discuss the accordance of the synthetic data generation strategy. Finally, Sec.~\ref{sec:conclusion} concludes the paper.

\section{Related work}
\label{sec:biblio}

The earliest speech bandwidth extension algorithms, usually applied to telephony applications, were performed with pure signal processing algorithms like spectral folding \cite{makhoul1979high}, Linear Predictive Coding \cite{chennoukh2001speech}, modulation techniques \cite{epps2000wideband,de2002yin} or non-linear processing \cite{iser2008bandwidth}. This simple procedure has also been used in the context of in-ear microphones \cite{bouserhal2017ear} with fair results, yet to be improved. This method creates missing harmonics in the high frequencies but cannot recover missing formants and fricatives. The earliest data-driven approaches have subsequently offered a more realistic extension. Those approaches are composed of several building blocks, including a statistical model that aims to estimate the high band spectral envelope. In many cases, this statistical model is one of the following: codebooks \cite{yoshida1994algorithm}, Gaussian Mixture Model \cite{park2000narrowband}, Hidden Markov Model \cite{jax2003artificial} and even some neural networks \cite{iser2003neural}. Although the quality is generally better with those methods, overly smoothed spectra are still produced at the expense of speech naturalness.

Recent advances in neural speech synthesis \cite{oord2016wavenet,shen2018natural,prenger2019waveglow,kong2020diffwave,yamamoto2020parallel,kong2020hifi} have proven that end-to-end deep learning is state-of-the-art in terms of simplicity and sample quality. Therefore, deep learning seems promising to accomplish this extreme bandwidth extension task. Indeed, the ability of neural networks to extract relevant features for the downstream task will allow the matching of high and low frequency contents. Raw waveform input is preferred over handcrafted features like spectrogram, mel-spectrogram \cite{davis1980comparison}, or Mel-frequency cepstral coefficients (MFCCs) \cite{bogert1963quefrency} to minimize human processing and let the network build its representation. This trend is endorsed by several works in the audio field \cite{dai2017very,germain2019speech,baevski2020wav2vec,goel2022s} and especially for bandwidth extension (synonym of audio super-resolution) to avoid rebuilding the phase separately \cite{ling2018waveform,birnbaum2019temporal,hao2020time,kuleshov2017audio}.

The use of raw audio can also be combined with multiband processing to speed up inference, as in \textit{DurIAN} \cite{yu2019durian}, \textit{MB-MelGAN} \cite{yang2021multi} or \textit{RAVE} \cite{caillon2021rave}. The speech signal is therefore processed at a reduced sampling rate thanks to the decomposition, unlike other super-resolution networks \cite{wang2021towards}, which use an input signal sampled at the target sampling frequency. To pursue the objective of fast inference, a fully convolutional architecture has been preferred like in \cite{oord2016wavenet,kuleshov2017audio}, specifically U-Net-like as other audio-to-audio tasks \cite{stoller2018wave,bosca2021dilated}. The upsampling layers use transposed convolutions \cite{zeiler2010deconvolutional} instead of subpixel layers \cite{shi2016real}. Transposed convolutions do not produce checkerboard artifacts when kernel size and strides are chosen to avoid overlapping disparities, as explained in \cite{odena2016deconvolution}.

In addition, a simple reconstruction loss may be insufficient for conditional generation, producing unrealistic samples. As shown in \cite{kumar2019melgan,kim2019bandwidth,kong2020hifi,eskimez2019adversarial}, adversarial networks \cite{goodfellow2020generative} can significantly improve the naturalness of the produced sound. Multiple discriminators are even used in \cite{hao2020time,tagliasacchi2020seanet,kim2019bandwidth} to focus signal discrimination at different time scales. Moreover, \textit{feature matching} is also encouraged for the reconstruction loss because it allows to enhance the produced sound quality in an end-to-end fashion. Indeed, discriminators' embeddings are excelling at building a relevant representation for our problem; it is therefore consistent to compute distance based on those features. Alternatives to this approach are either the $L_1$/$L_2$ norms in the time domain, which are known to be misaligned with human perception, or complex losses like multiscale Short Time Fourier Transform, which depend on chosen hyperparameters \cite{feng2019learning,kuleshov2017audio}, thus increasing tuning efforts.

Regarding the specific literature on blind (or non-multi modal) speech enhancement for BCMs, different approaches adopted classic processing to achieve bandwidth extension \cite{shahina2007mapping,shin2012survey,turan2013enhancement,bouserhal2017ear}. Subjective quality evaluations have proven those approaches to be inadequate for this task.
Then, neural networks started to be employed, firstly as a processing block among others \cite{park2019speech} to estimate an enhancement function in a fixed feature domain combined with time-domain filtering.  Subsequent research then began to carry out the improvement task and ultimately to perform the enhancement as end-to-end tasks. Among published manuscripts on the subject, the works of Yuang Li et al.\cite{li2022enabling}, Hung-Ping Liu et al.\cite{liu2018bone}, and Dongjing Shan et al.\cite{shan2018novel} applied this approach for bone conduction microphones,  and Mattes Ohlenbusch et al. \cite{ohlenbusch2022training} to in-ear microphones. The main drawback of those approaches relies on the fact that they are based on a pure reconstructive loss, eventually with a regularization part. As they expressed in their articles, \textit{an audible difference between the target and the processed signals remains}. This statement may be irrevocable due to the limited information left in the signal captured by BCMs. However, GANs \cite{goodfellow2014generative} can produce realistic signals by slightly deviating from the reference. The task of speech enhancement for speech capture with body-conduction microphones is thus complicated. Indeed, \cite{li2014multisensory,tagliasacchi2020seanet,wang2022multi,wang2022fusing} only used BCMs as a conditional signal for enhancement in a multi-modal framework. Moreover, even if BCMs mainly capture speech, residues of external and physiological noises persist \cite{bouserhal2018classification} and would necessitate denoising. Hopefully, deep learning models can perform the denoising task simultaneously with the bandwidth extension. \cite{bouserhal2017ear} has also proven that the contaminating noise knowledge was helpful, although we will not capitalise on this particular knowledge in the present paper.

Lastly, this research area lacks large public corpora that use body-conduction microphones to reliably train deep models. The ABCS \cite{wang2022abcs} and EMSB\footnote{\url{https://github.com/elevoctech/ESMB-corpus}} datasets, which consist of either bone or in-ear and air-conducted Mandarin speech pairs, are currently the largest corpora, consisting of 42 hours and 128 hours of speech, respectively. Another smaller public\footnote{freely available upon request from a research institution} dataset is Speech in-EAR (SpEAR) database proposed in \cite{bouserhal2019ear} with 25 participants, split in French/English speakers. Other private datasets emerged, like \cite{li2022enabling}, which introduced 200 minutes of speech recorded via bone conduction. The dataset was large enough to train on, likely due to their model's meager number of parameters (4.5k for the lightest model). \cite{ohlenbusch2022training} opted for a different strategy with their overall 30 minutes in-ear captured speech. The limited-size dataset was first used to simulate meaningful degradations, taking into account the body-produced noise used to train their model. Finally, they re-used real data to fine-tune their model's decoder.

\section{In-ear microphone study}
\label{sec:micro}

The selected BCM is an early prototype based on a MEMS microphone (STM MP34DT01) driven by an STM32 H7 microcontroller \cite{bionear} developed by the ISL and Cotral Lab. This device takes advantage of the speaker's hearing protection by being placed inside a custom-molded earplug, which increases communication capabilities in challenging and noisy environments. The reference speech signals are captured by a B$\&$K Type 4192 condenser microphone connected to a TEAC LX10 data recorder. The reference and in-ear signals are recorded at 48 kHz, resampled at 16 kHz and finally synchronized using cross-correlation. We collected utterances of the Combescure's phonetically balanced sentences \cite{combescure198120} from a single speaker.

\subsection{In-noise comparison with traditional microphone}
\label{sec:innnoisecomparison}
\begin{figure}[htb]
  \centering
  \centerline{\includegraphics[width=0.9\linewidth]{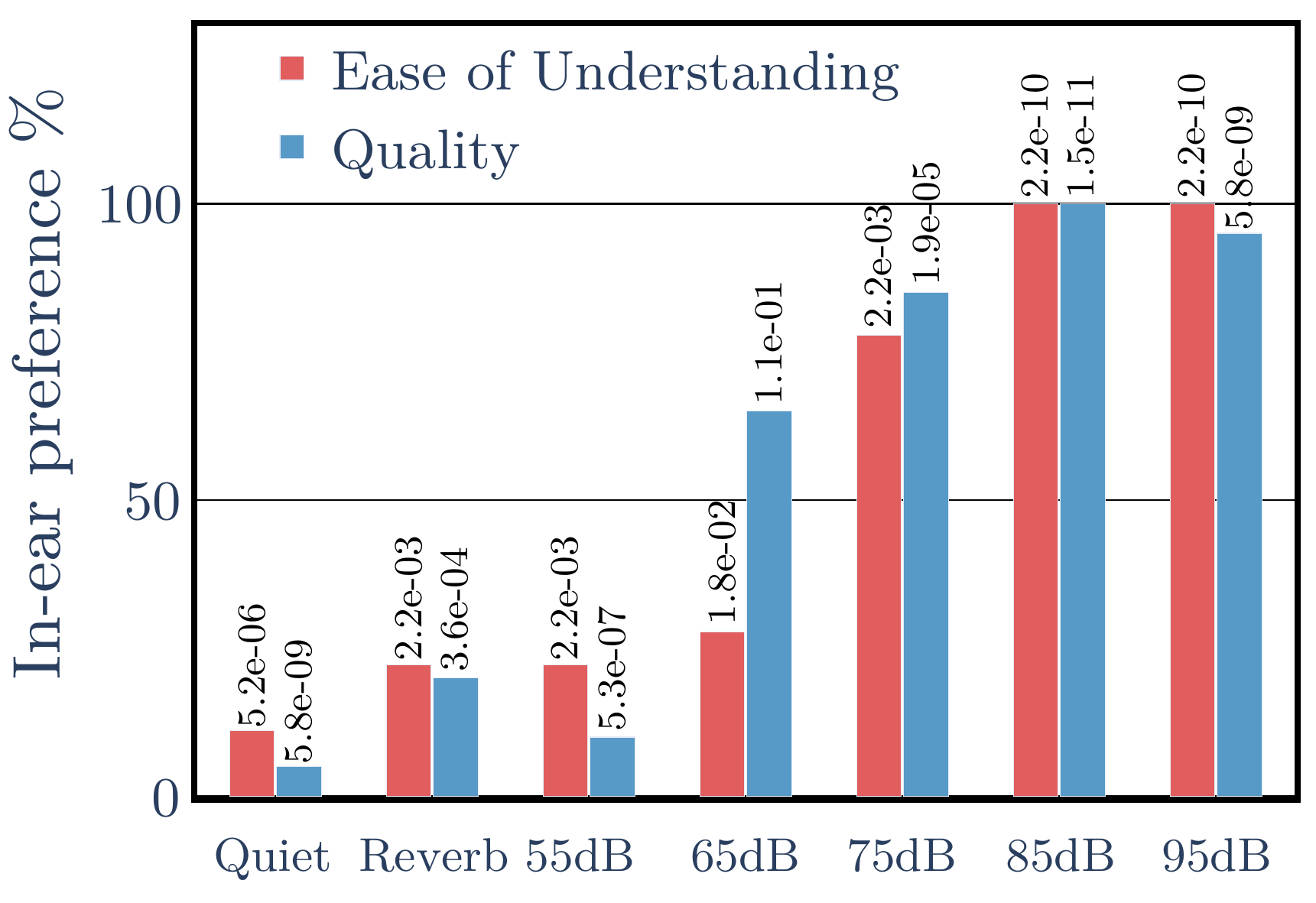}}
\caption{A/B testing results: in-ear vs traditional microphones. The p-values shown at the top of each bar indicate the significance of the preferred microphone.}
\label{fig:miccomparison}
\end{figure}

This section aims at justifying that BCMs are more suitable for noisy environments and at establishing a rough estimate of the noise level above which their use should prevail. We conducted subjective A/B preference tests to compare our in-ear microphone with a traditional microphone. A single speaker was recorded simultaneously by both microphones in different acoustic environments, ranging from a quiet acoustically treated room to a reverberating room with several levels of surrounding noise. When surrounding noise is present, the speaker naturally produces Lombard speech \cite{brumm2011evolution}. Comparisons are performed using 7 different utterances from the Combescure's sentences \cite{combescure198120},  which have been recorded in an audiometric booth (IAC Acoustics and walls covered with acoustic foam), and in a reverberating room with pink noise levels $\{$ $\emptyset$ ,55dB, 65dB, 75dB, 85dB, 95dB $\}$ without any enhancement techniques. We recruited 38 participants and used the GoListen platform \cite{barry2021go} to conduct the test. Participants were asked whether they preferred in-ear or classic recordings for all acoustic environments. Participants were divided into two groups to rate either the quality or the ease of understanding of the audio samples. Obtained results are presented in Fig.~\ref{fig:miccomparison} along with the corresponding p-values for each A/B comparisons.

According to Fisher's exact test and a significance rate of 1\%, the obtained results allow us to conclude that the use of an in-ear microphone is preferred for both ease of speech understanding and sound quality for surrounding noise levels of 75 dB or more. On the other hand, a traditional microphone is endorsed for ease of understanding and quality for noise levels below or equal to 55dB. No statistically significant difference can be drawn for a 65dB noise level.

\subsection{Degradation study}
\label{sec:degradation}

In-ear own voice capture is more adapted for applications in noisy environments because it mainly contains speech without external noise. However, the acoustic wave propagation between the vocal tract and the transducers causes irreversible information loss: almost no relevant speech signal is picked up above a threshold frequency. Complex interactions with tissues are also responsible for phase shifts and anti-resonances.

This phenomenon is further influenced by the occlusion effect \cite{brummund2014three} due to the fitting of the individual protectors, causing speech to resonate inside the ear canal. This aspect causes an amplification of the remaining signal, leading the wearer to hear an amplified version of their own voice. The occlusion effect is therefore the consequence of wearing an earplug, but it is also necessary in order to obtain an in-ear signal that is not significantly degraded by environmental noise.

A first coarse approximation of those degradations can be modeled by a linear impulse response $\psi$ that allows to estimate the in-ear signal $x$ from the emitted signal $y$ :

 \begin{equation}
 x(t)= (\psi*y)(t)
\end{equation}
To evaluate the corresponding transfer function, we simultaneously use the in-ear prototype and a regular microphone placed in front of the speaker's mouth under noise-free conditions. The absence of noise allows us to consider airborne speech as the emitted signal. The degradation filter estimates $\{\tilde{\Psi}_i\}_{i \in [1,53]}$ were obtained with cross power spectral densities $\{P_{yx,i}\}_{i \in [1,53]}$ and $\{P_{yy,i}\}_{i \in [1,53]}$ approximated by Welch's method \cite{welch1967use}, Eq.~\ref{eq:hf}. Short Time Fourier Transforms were computed on $512$ samples corresponding to $32$~ms for the $16$~kHz sampling rate used during this analysis. Welch's method has a temporal horizon of $1.024$~second with a recovery rate of $50\%$. A Voice Activity Detection (VAD) pre-processing based on a simple reference's power thresholding was applied to select meaningful segments. The reference and in-ear signals were normalized before calculating the cross power spectral densities because the in-ear microphone is not calibrated. Therefore, the shape of the transfer function is correct as a function of frequency, but the absolute amplitude does not reflect differences in sound pressure.

\begin{equation}
 \tilde{\Psi}_i(f)=\frac{P_{yx,i}(f)}{P_{yy,i}(f)}~, \forall i \in [1,53]
 \label{eq:hf}
\end{equation}

\begin{figure}[htb]
  \centering
  \centerline{\includegraphics[width=0.9\linewidth]{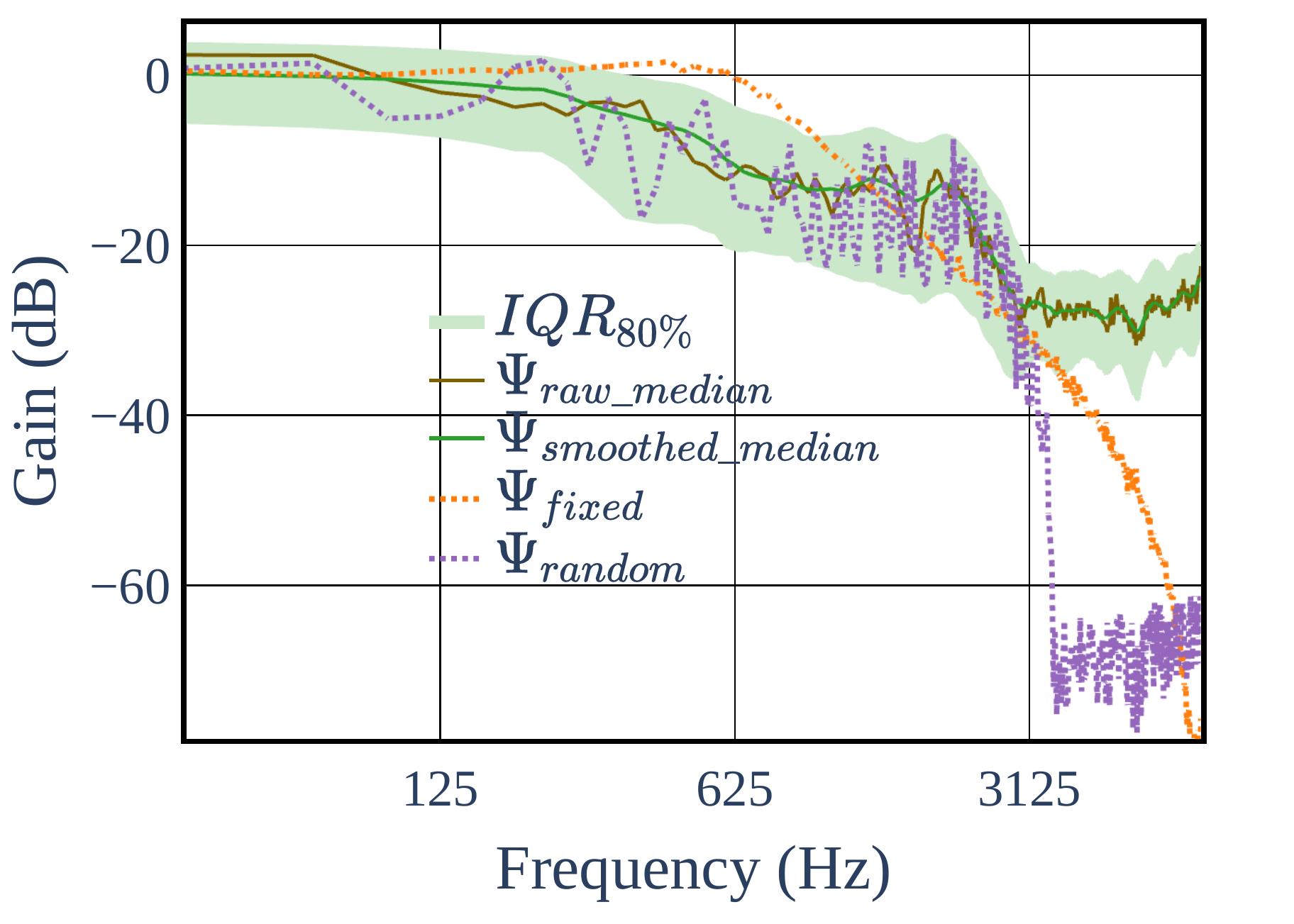}}
\caption{Transfer function of the in-ear transducer}
\label{fig:degradation}
\end{figure}

$53$ estimates were necessary to produce a robust estimation of the transfer function, noted $\Psi_{raw\_median}=median(\{\tilde{\Psi}_i\}_{i \in [1,53]})$, because speech signals are not stationary. The analysis was performed on a single-person recording of $23$~seconds after the VAD processing, corresponding to 10 utterances of Combescure's sentences \cite{combescure198120}. As $\Psi_{raw\_median}$ is still noisy we performed a smoothing step to obtain $\Psi_{smoothed\_median}$ which is plotted in Fig.~\ref{fig:degradation}, surrounded by its $10\%$ and $90\%$ percentiles, illustrated by $IQR_{80\%}$.

\begin{figure}[htb]
  \centering
  \centerline{\includegraphics[width=0.9\linewidth]{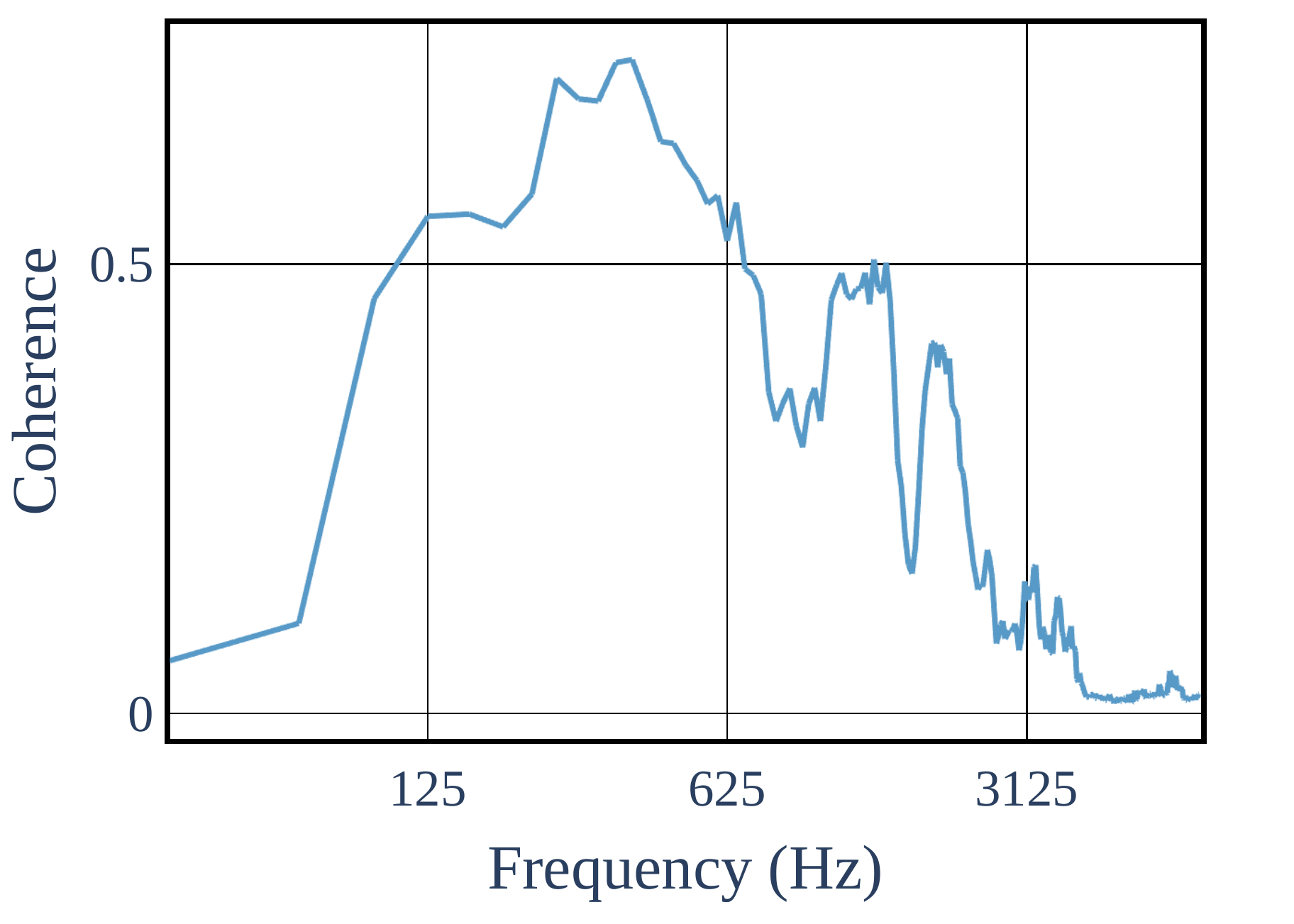}}
\caption{Coherence function of the in-ear transducer}
\label{fig:coherence}
\end{figure}

The estimated coherence function $\tilde{C}_{yx}$, defined on Eq.~\ref{eq:coherence} and represented on Fig.~\ref{fig:coherence}, highlights an absence of causality between $x$ and $y$ above 3kHz. Hence, Fig.~\ref{fig:degradation} does not make sense above that frequency.

\begin{equation}
 \tilde{C}_{yx}(f)=\frac{|P_{yx}(f)|^2}{P_{xx}(f)P_{yy}(f)}
 \label{eq:coherence}
\end{equation}

\begin{figure}[hb!]
  \centering
  \centerline{\includegraphics[width=0.9\linewidth]{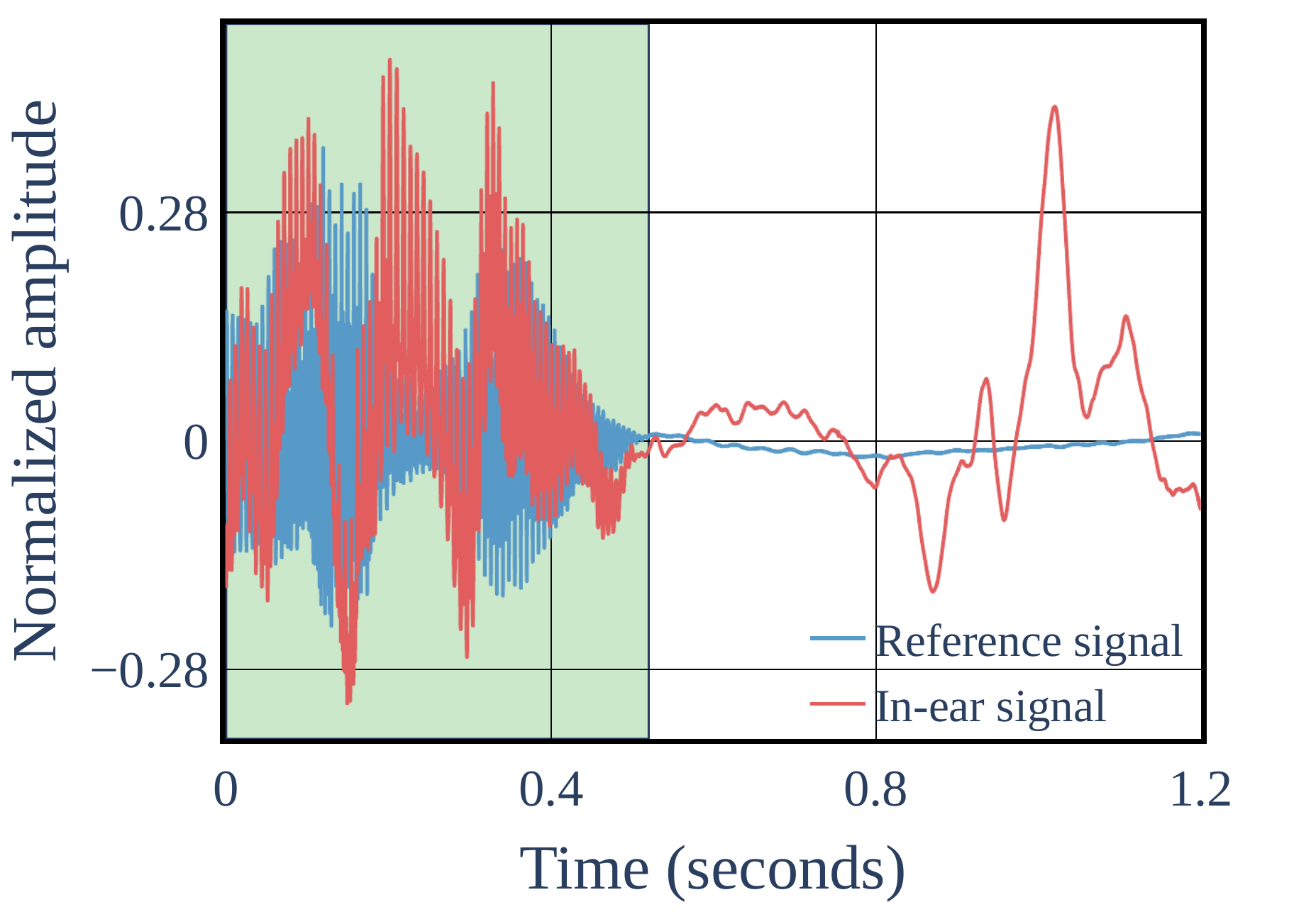}}
\caption{Time domain representation of speech signals captured in a quiet environment. Active speech is presented in green area.}
\label{fig:temporal}
\end{figure}

This shows that the in-ear BCM allows to only capture relevant speech content inside the ear canal for frequencies $\{f~|~\Psi_{smoothed\_median}(f)>-20dB,\forall f\in \mathbb{R}^+\}$ \textit{i.e.}~in a range below 2 kHz. Indeed, Fig.~\ref{fig:degradation} indicates that the in-ear microphone exhibits a very high attenuation at mid and high frequencies: no relevant signal is present in this frequency range. Interestingly, at very low frequencies \textit{i.e.}~~below 80 Hz, the coherence function in Fig.~\ref{fig:coherence} is also close to zero, which denotes a poor correspondance between the two signals. The physiological sounds (\textit{e.g.}~ swallowing, blood flow, tongue clicking, teeth grinding) are responsible for this phenomenon, as they are only sensed by the in-ear transducers. Some additional phenomenons like microphonics and movement artifacts may also occur. A time domain representation of the synchronized capture in Fig.~\ref{fig:temporal} highlights this difference in the quiet region, for $t>0.5$~s. For the $[80, 300]$ Hz range, the occlusion effect and small shifts in formant frequencies may occur.

Finally, two anti-resonances are observed in Fig.~\ref{fig:coherence} at 900 Hz and 1700 Hz, corresponding to vibration nodes of the occluded ear canal and propagation in the bones and tissues of our subject. It is noteworthy that those observations are not universal: acoustic paths differ among speakers because their bone and biological tissue structures are unique - as well as their ear canal geometries and properties - which results in different spectral properties.

\subsection{Simulation of the dataset}
\label{sec:simulation}

Deep learning-based approaches are only efficient in large data regimes; the few minutes of in-ear samples currently available to us are highly insufficient for supervised training. Therefore, we adopted a simulated corrupted wideband speech from the French Librispeech dataset \cite{pratap2020mls} in an in-ear-like fashion, along with a data augmentation strategy. In the present paper, we simulated two kind of transfer functions to filter the clean speech data: $\Psi_{fixed}$ and $\Psi_{random}$ which are jointly plot on Fig.~\ref{fig:degradation}.

In both cases, a gaussian white noise with a power -23 dB below the low-pass filtered signal is added. This noise intends to play the role of physiological noise. It is also masking any high frequency residues.

\underline{$\Psi_{fixed}$:} This fixed degradation, used in Sec.~\ref{sec:experiments}, is obtained using an autoregressive moving-average model. $\Psi_{fixed}$ is a 2\textsuperscript{nd} order low-pass filter with a cutoff frequency of $600$~Hz and unitary Q-factor that is applied using a \emph{filtfilt}\footnote{consists in applying a digital filter forward and backward to a signal.} procedure to ensure zero phase shift.

\underline{$\Psi_{random}$:} This ever-changing degradation is constructed to fall within the green area of Fig.~\ref{fig:degradation}. It is more realistic as it fits better to $\Psi_{smoothed\_median}$ and has some randomness. Indeed, $\Psi_{random}$ is sampled from a log-uniform distribution with $IQR_{80\%}$ bounds and brought to a very low gain above 3kHz with an Hann apodization function. We will use this transfer function in Sec.~\ref{sec:phirandom} to discuss the limitations of the linear time-invariant modeling of the in-ear degradation.

Those simulated degradations might lack some realism but still ensure a wide application field for developed algorithms and the ability to focus on the bandwidth extension issue. Naturally, this simulation approach would involve minimizing the disparity between the simulated and real data; however, this process is highly time-consuming and does not guarantee improved performance. Addressing this discrepancy would require incorporating additional speakers, surpassing our current assumption of a linear impulse response, and adopting complex physical models, such as in \cite{blonde2023numerical}. Moreover, it would necessitate the realistic blending of pre-recorded physiological noises with speech signals. Instead, we have opted for a simpler yet adequately challenging degradation approach to compare various methods. In parallel, we are focusing on collecting a substantial dataset of speech capture with different BCMs for a subsequent study. Once this dataset is complete, the step of enhancing simulation relevance will be bypassed.

\section{EBEN}
\label{sec:eben}

\subsection{Theory of Pseudo Quadrature Mirror Filter}
\label{sec:pqmf}

\begin{figure}[htb]
  \centering
  \centerline{\includegraphics[width=0.9\linewidth]{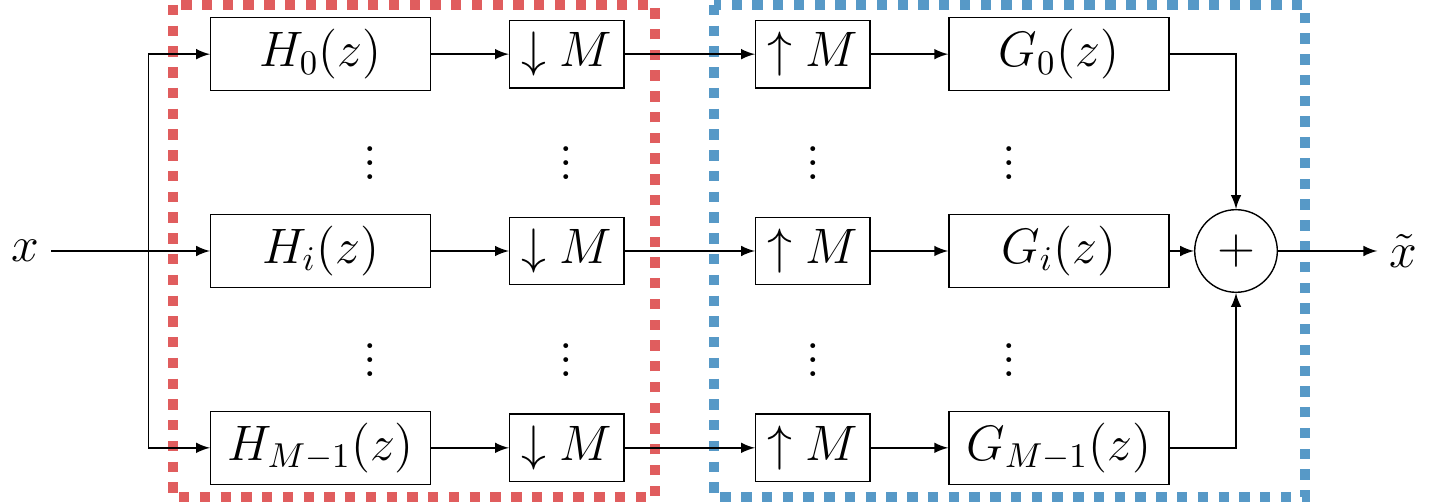}}
\caption{PQMF Analysis and Synthesis : block-diagram}
\label{fig:aands}
\end{figure}

\begin{figure}[htb]
  \centering
  \centerline{\includegraphics[width=0.9\linewidth]{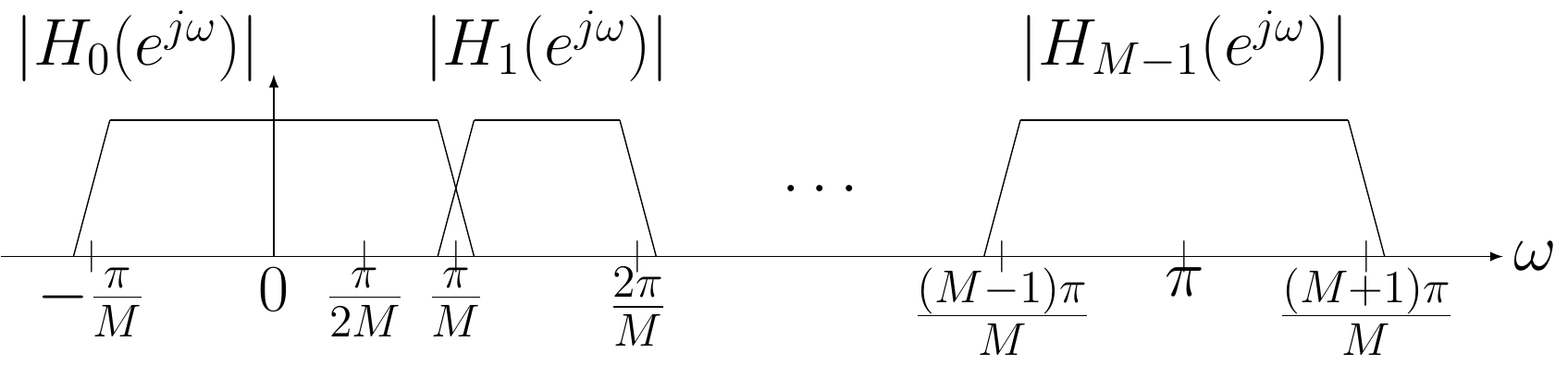}}
\caption{Frequency response of a PQMF ﬁlter bank}
\label{fig:bode}
\end{figure}

The Quadrature Mirror Filter (QMF) banks, introduced in \cite{rothweiler1983polyphase}, are a set of analysis filters $\{H_i\}_{i \in [0,M-1]}$ used to decompose a signal into several non-overlapping channels of same bandwidth, and synthesis filters $\{G_i\}_{i \in [0,M-1]}$ used to recompose the signal afterward. Fig.~\ref{fig:aands} shows the entire pipeline. Those filters are obtained from frequency translations of the same low-pass prototype filter $h[n]=\mathcal{Z}^{-1}\{H(z)\}$. A typical frequency response for an M-band Pseudo-QMF (PQMF) bank is given in Fig.~\ref{fig:bode}.

The reconstruction is exact if $\{H_i\}_{i \in [0,M-1]}$ and $\{G_i\}_{i \in [0,M-1]}$ have an infinite support. In practice, this is impossible, but Truong Nguyen proposed a near-perfect reconstruction in \cite{nguyen1994near} by constraining the prototype filter to be a linear-phase spectral factor of a $2Mth$ band filter, significantly reducing aliasing. In other words, the analysis and synthesis impulse responses noted respectively $h_i[n]=\mathcal{Z}^{-1}\{H_i(z)\}$ and $g_i[n]=\mathcal{Z}^{-1}\{G_i(z)\}$, are given by Eq.~\ref{eq:impulses} where $N$ is the filter length.

\begin{equation}
\begin{split}
  \left\{
    \begin{array}{ll}
        h_i[n] = 2h[n] \cos \left( (2i+1)\frac{\pi}{2M}  \left( n - \frac{N-1}{2} \right) + (-1)^{i}\frac{\pi}{4} \right) \\
        g_i[n] = 2h[n] \cos \left( (2i+1)\frac{\pi}{2M}  \left( n - \frac{N-1}{2} \right) - (-1)^{i}\frac{\pi}{4} \right)
    \end{array}
\right. \\
,~~ 0\leq n \leq N-1,~~ 0\leq i \leq M-1
\end{split}
\label{eq:impulses}
\end{equation}

Then, Yuan-Pei Lin and PP Vaidyanathan \cite{lin1998kaiser} proposed a more straightforward design methodology by constructing the prototype from a Kaiser window and filling the following conditions:
\begin{itemize}
 \item Make the prototype filter close to zero out of its passband to minimize the aliasing.
\begin{equation}
|H(e^{j \omega})| \approx 0~~~ \mbox{for}~ |\omega|>\frac{\pi}{M}
\end{equation}
 \item Make the prototype filter close to one into its passband to minimize the distorsion.
 \begin{equation}
|H(e^{j \omega})| \approx 1~~~\mbox{for}~ |\omega|<\frac{\pi}{M}
\end{equation}
\end{itemize}

Given the desired stopband attenuation and transition bandwidth, these requirements directly translate into a one-degree-of-freedom optimization criterion at the prototype's cutoff frequency. This criterion is minimized to find the optimal cutoff frequency for some $M$ and $N$. Also note that although minimal band overlap implies very low reconstruction error, it is not equivalent; in fact, phase opposition phenomena between the bands also contribute to the elimination of redundant content in the synthesis phase. In practice, a kernel size of $N=8M$ is sufficient for pseudo-perfect reconstruction (signal-to-error ratio of $55$ dB), and a kernel size of $N=128M$ is sufficient for pseudo-perfect separation of frequency content between bands. In this article, we have used a convolution kernel of $N=8M$, which is fast to compute and sufficient to separate frequency content.

\subsection{Model architecture}
\label{sec:architecture}

\subsubsection{Generator}
\begin{figure*}[ht!]
  \centering
    \begin{subfigure}[t]{\textwidth}
        \centering
        \centerline{\includegraphics[width=\textwidth]{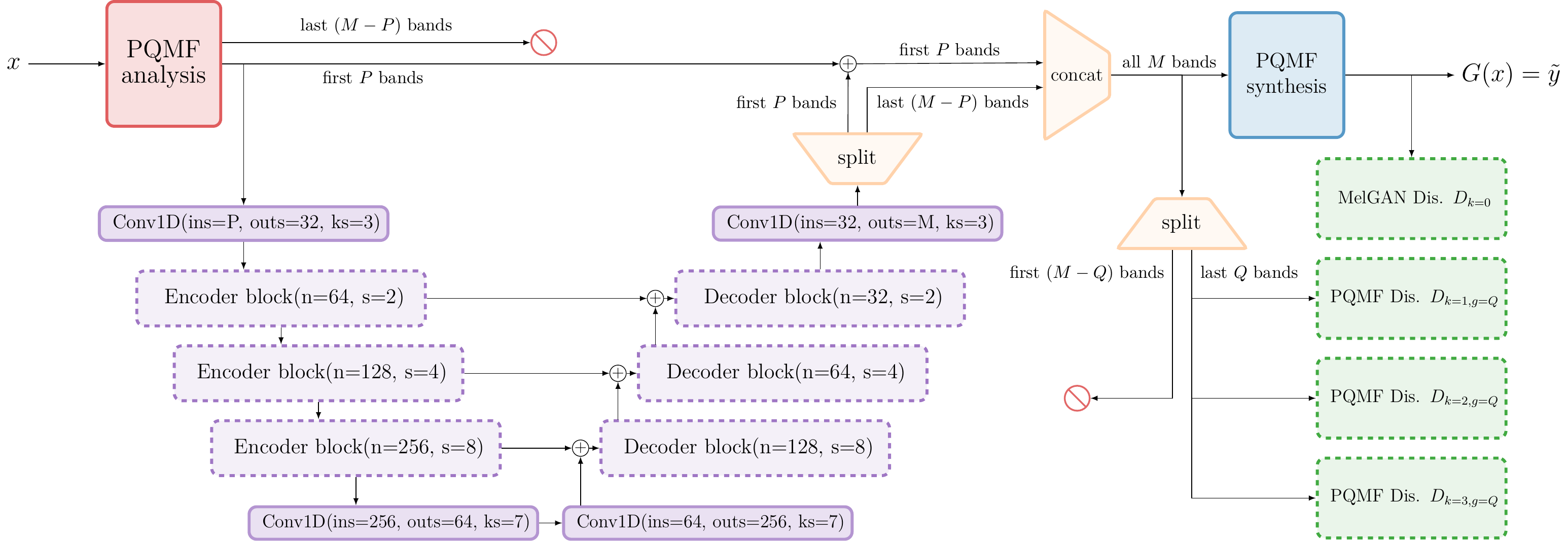}}
        \caption{Overall architecture}
        \label{fig:overview}
        \vspace{1cm}
    \end{subfigure}
    \begin{subfigure}[t]{0.333\textwidth}
        \centering
        \includegraphics[width=0.96\textwidth]{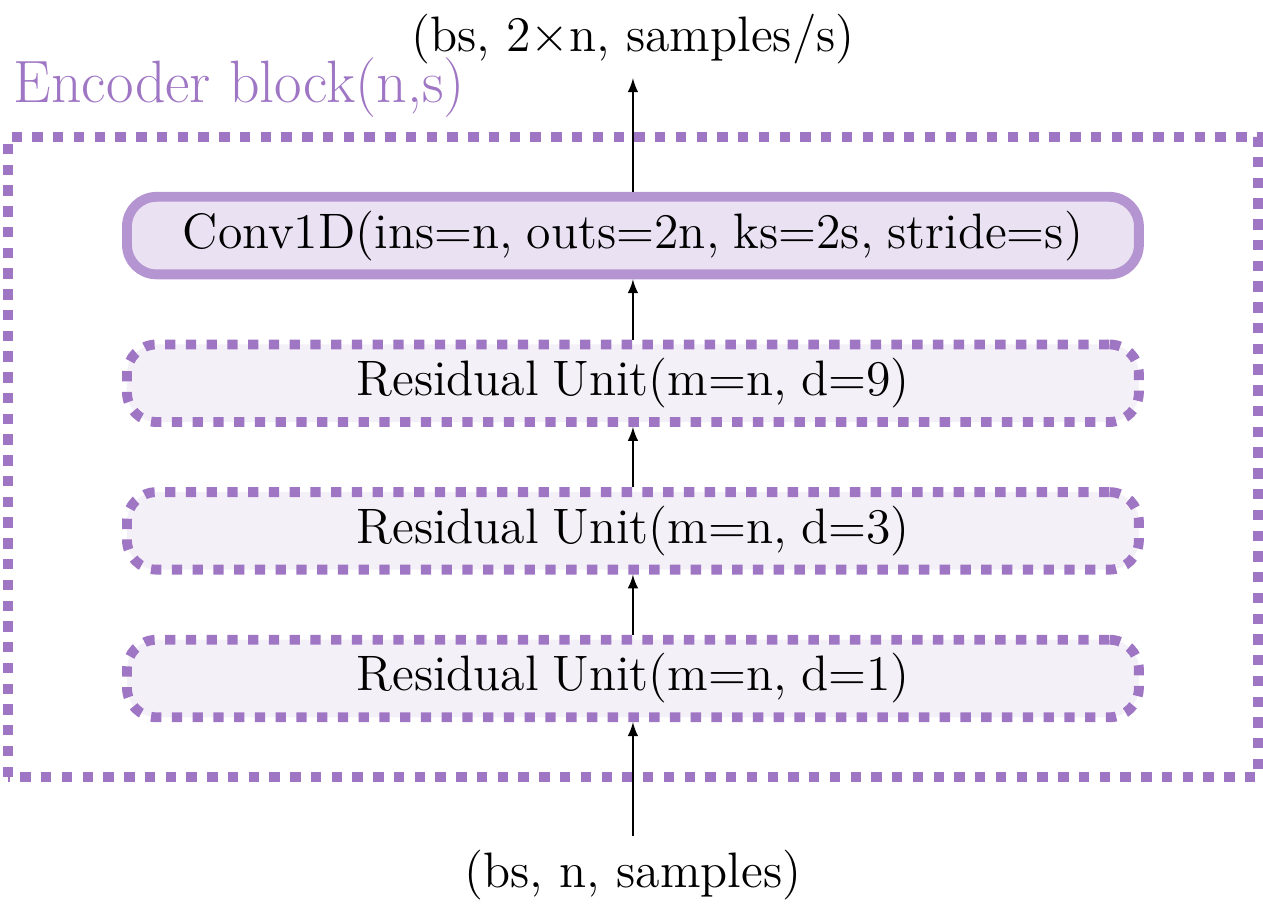}
        \caption{Encoder block}
        \label{fig:enc}
    \end{subfigure}%
    \begin{subfigure}[t]{0.333\textwidth}
        \centering
        \includegraphics[width=0.96\textwidth]{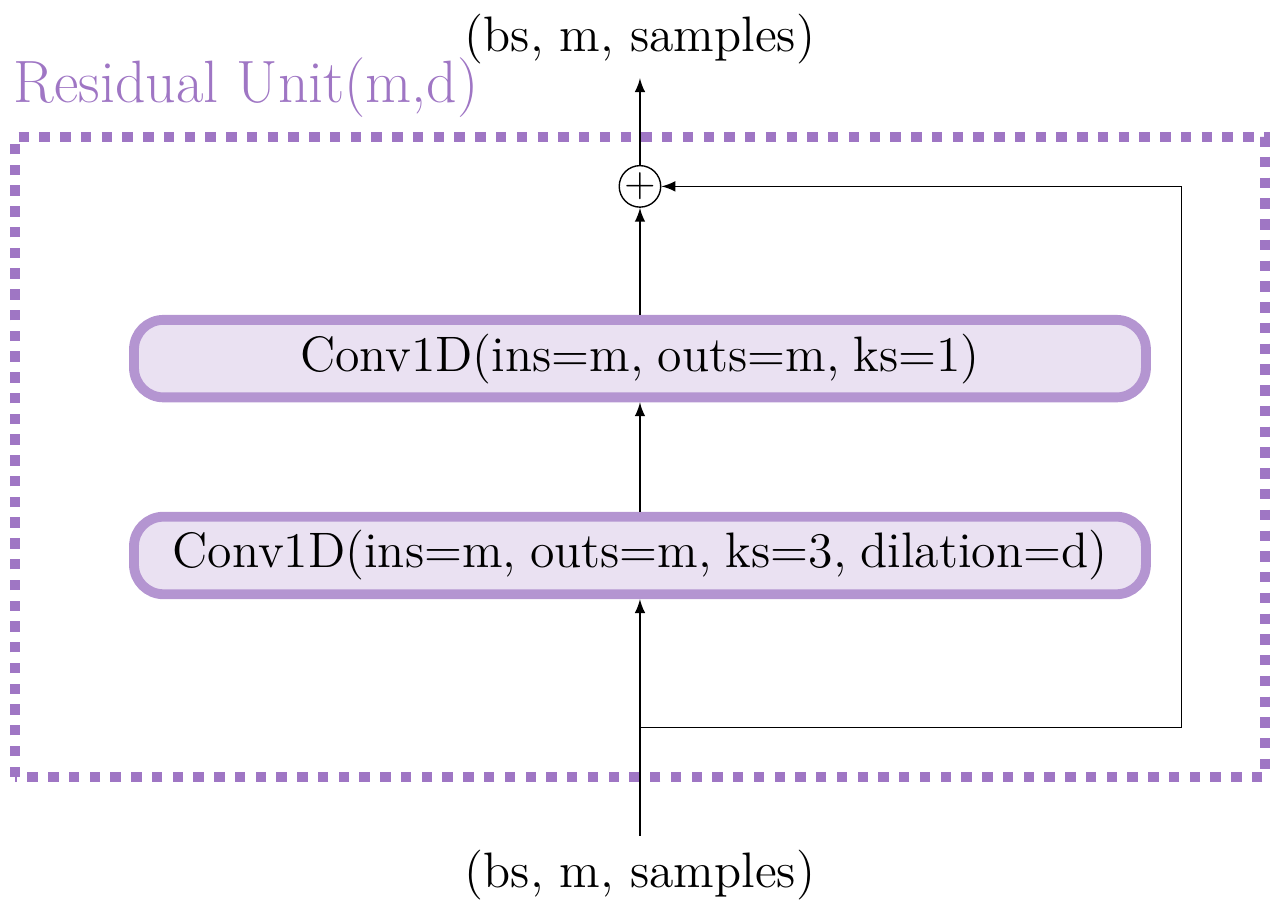}
        \caption{Residual Unit}
        \label{fig:resi}
    \end{subfigure}%
    \begin{subfigure}[t]{0.333\textwidth}
        \centering
        \includegraphics[width=0.96\textwidth]{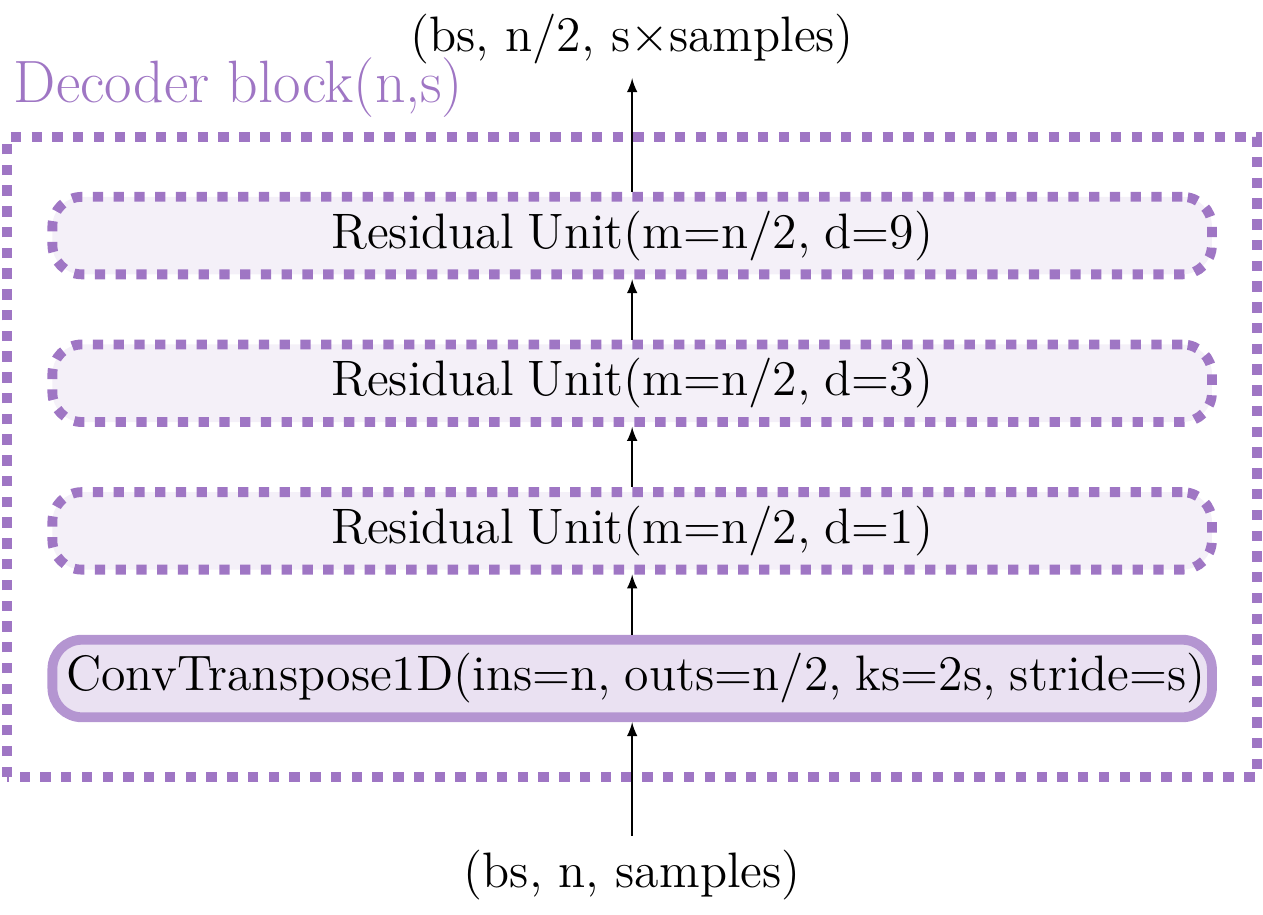}
        \caption{Decoder block}
        \label{fig:dec}
        \vspace{1cm}
    \end{subfigure}
    \begin{subfigure}[t]{0.5\textwidth}
        \centering
        \includegraphics[width=0.95\textwidth]{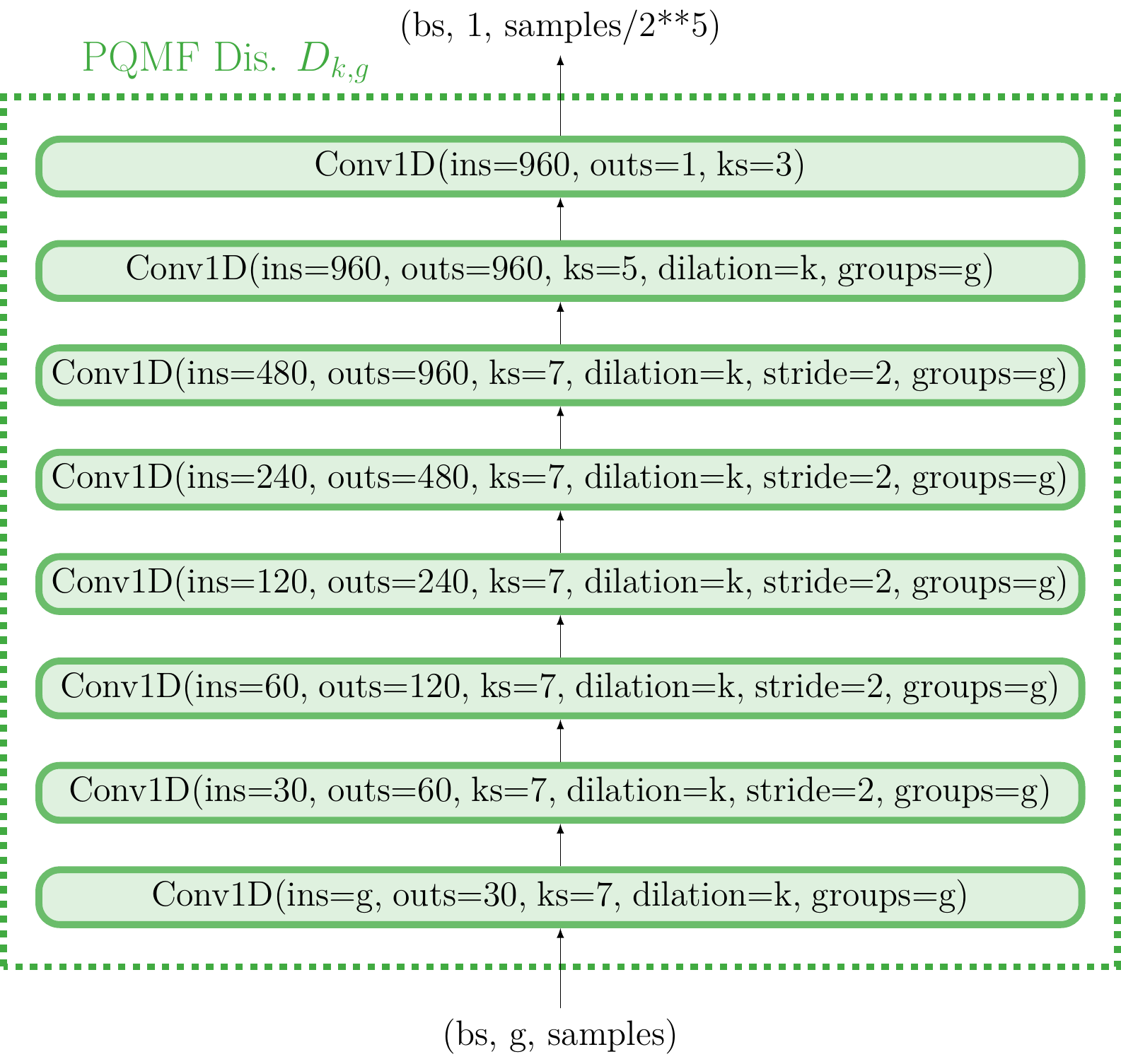}
        \caption{PQMF discriminator}
        \label{fig:dpqmf}
    \end{subfigure}%
    \begin{subfigure}[t]{0.5\textwidth}
        \centering
        \includegraphics[width=0.95\textwidth]{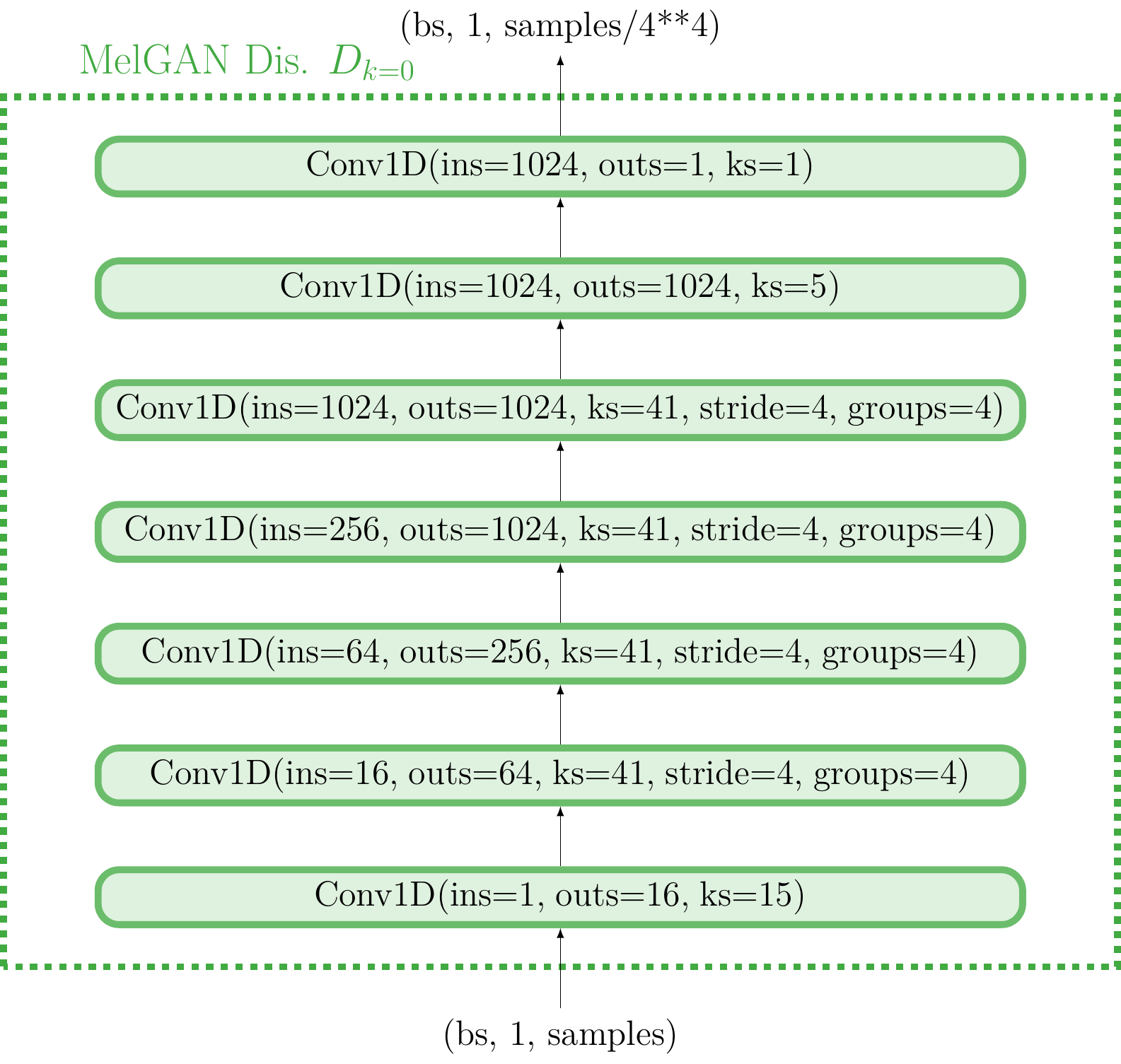}
        \caption{MelGAN discriminator}
        \label{fig:dmelgan}
        \vspace{0.5cm}
    \end{subfigure}
\caption{Architecture of EBEN. \textit{ins}: input channels. \textit{outs:} output channels. \textit{ks:} kernel size}
\label{tikz:overall}
\end{figure*}

Unlike frequency approaches \cite{kumar2019melgan,kong2020hifi,lagrange2020bandwidth}, which require massive 2D convolutional operations to extract meaningful features from spectrograms or heavy waveform approaches \cite{kuleshov2017audio,tagliasacchi2020seanet,li2021real,wang2021towards,li2022enabling,su2021bandwidth} which directly process the audio at the targetted sampling rate, we propose for EBEN to encapsulate a lightweight U-Net-like generator between a PQMF analysis layer and a PQMF synthesis layer.
This enclosure reduces the model's memory footprint by decreasing the first embedding sample rate by a factor of $M$. It also makes it possible to keep only $P$ subbands with voice content to feed to the first convolution and the last convolution via the most external skip connection. $P$ must lie between $1$ and $M$. Moreover, the number of encoder/decoder blocks is reduced to meet the constraints of real-time applications. Global architecture is exhibited in Fig.~\ref{fig:overview} and subblocks in Fig.~\ref{fig:enc},\ref{fig:resi},\ref{fig:dec}. Convolutions are intertwined with Leaky ReLU activation functions with a negative slope of $0.01$. The last non-linearity in the generator is a Hyperbolic tangent placed right before the PQMF synthesis block, in order to bring back values between -1 and 1. Skip connections are additive. We also apply weight normalization \cite{salimans2016weight} on top of every convolution block with trainable weights, in order to ensure a fast convergence during training. Altogether, the EBEN generator is configured by $M$: the number of PQMF bands and $P$: the number of bands from which information is extracted.

\subsubsection{Discriminators}
EBEN's discriminators directly exploit the PQMF subbands as inputs without recombining nor upsampling the reconstructed subband signals. We adopt a multiscale ensemble discriminator approach, inspired by the work of Kumar \emph{et al.} in \cite{kumar2019melgan}, whose inputs are the $Q$ upper bands of the PQMF decomposition, similarly to \cite{mustafa2021stylemelgan}. Due to the divisibility constraint on the number of input and output channels by the number of groups, $Q$ must be one of $\{1,2,3,5,6,10,15\}$. Like $P$, it must also satisfy $1\leq Q\leq M$. The ensemble of discriminators analyzes the generated subband signals at different time scales and helps to improve their quality via the adversarial process, even though each discriminator is relatively simple. The subband discriminators $\{D_k\}_{k\in[1,2,3]}$ exhibit similar receptive fields to the original multiscale MelGAN discriminators \cite{kumar2019melgan}. Moreover, we combined our PQMF discriminators with the full scale MelGAN discriminator $D_{k=0}$ to ensure coherence between bands. The exact architecture of discriminators are displayed in Fig.~\ref{fig:dpqmf} and Fig.~\ref{fig:dmelgan} together with their positioning in the overall system Fig.~\ref{fig:overview}. We kept Leaky ReLU as an activation function but used a stronger negative slope of $0.2$ to allow for a better gradient transmission to the generator. We also maintained the weight normalization technique. Overall, the EBEN discrimators are configured by $M$: the number of PQMF bands and $Q$: the number of enhanced subbands.

\subsection{Loss functions}
\label{sec:loss}

At each batch, we train alternatively the ensemble of discriminators $\{D_k\}_{k\in[0,1,2,3]}$ to minimize $\mathcal{L_D}$ defined on Eq.~\ref{eq:ld} and the generator $G$ to minimize  $\mathcal{L_G} = \mathcal{L_G}^{adv} + 100 \times \mathcal{L_G}^{rec}$ where $\mathcal{L_G}^{adv}$ and $\mathcal{L_G}^{rec}$ are respectively defined on Eq.~\ref{eq:lga} and Eq.~\ref{eq:lgr}.
Our loss setup is inspired by \cite {tagliasacchi2020seanet}: $\mathcal{L_D}$ and $\mathcal{L_G}^{adv}$ are a classical hinge loss while $\mathcal{L_G}^{rec}$ is a feature matching loss. Using discriminators embeddings for the reconstructive loss allows focusing on the semantic of the signal, which is harder to operate in the time domain because useful information is drowned out amid useless details.

In the underneath definitions, $D_{k,t}^{(l)}$ represents the layer $l$ of the discriminator (among $L_k$ layers) of scale $k$ (among $K$ scales) at time $t$. $F_{k,l}$ and $T_{k,l}$ are the number of features and temporal length for given indices. We kept $x$ for in-ear signal and $y$ for the reference.

\begin{equation}
\begin{split}
 \mathcal{L_D}= E_y\left[ \frac{1}{K} \sum_{k \in [0,3]} \frac{1}{T_{k,L_k}} \sum_t max(0,1-D_{k,t}(y))\right] + \\ E_x\left[ \frac{1}{K} \sum_{k \in [0,3]} \frac{1}{T_{k,L_k}} \sum_t max(0,1+D_{k,t}(G(x)))\right]
 \label{eq:ld}
 \end{split}
\end{equation}

\begin{equation}
\mathcal{L}_\mathcal{G}^{adv}= E_x\left[ \frac{1}{K} \displaystyle \sum_{k \in [0,3]} \frac{1}{T_{k,L_k}} \sum_t max(0,1-D_{k,t}(G(x)))\right]
\label{eq:lga}
\end{equation}

\begin{equation}
\mathcal{L}_\mathcal{G}^{rec}= E_x\left[ \frac{1}{K} \displaystyle \sum_{\substack{k \in [0,3] \\ l \in [1,L_k [ }} \frac{1}{T_{k,l}F_{k,l}} \displaystyle \sum_t \| D_{k,t}^{(l)}(y)-D_{k,t}^{(l)}(G(x))\| _{L_1}  \right]
\label{eq:lgr}
\end{equation}

The generator's loss combination allows to generate audio samples as close as possible to the reference signal thanks to $\mathcal{L}_\mathcal{G}^{rec}$, while remaining creative at high frequencies when no information is available in the degraded signal (especially for fricatives) thanks to $\mathcal{L}_\mathcal{G}^{adv}$.

\section{Experiments and evaluation}
\label{sec:experiments}

\subsection{Training strategy}
We trained different models \cite{kuleshov2017audio,kong2020hifi,tagliasacchi2020seanet,li2021real} and the proposed EBEN model on the French Librispeech \cite{pratap2020mls} dataset resampled uniformly at 16kHz to reverse the $\Psi_{fixed}$ degradation applied on the fly. All the experiments were performed for two days on a single RTX 2080 Ti GPU with a $16$ batch size of 2-second randomly sliced audio, corresponding to 13 epochs for the EBEN model. Losses are optimized with Adam \cite{kingma2014adam} using a constant learning rate of $3\mathrm{e}{-4}$ and $\beta=(0.5,0.9)$ for EBEN and optimizers parameter values found in original papers for the other approaches. No parameter tuning nor early stopping was performed.
The EBEN set of hyperparameters is given by $\{M=4, P=1, Q=3\}$. We use $M=4$ here because this coarse slicing of the spectra is sufficient to separate frequency bands containing valuable cues from non-relevant frequency bands by taking $P=1$. Such a reduced number of frequency bands also allows to reduce the length of the PQMF kernel for the analysis and synthesis stages. The value $Q=3$ was also chosen because we assume that the first frequency band does not require significant enhancement with the proposed degradation.

\subsection{Objective evaluation}
\label{sec:objective}

\subsubsection{Speech quality metrics}
To evaluate the model performances, Tab.~\ref{tab:objective} highlights several objective metrics: Perceptual Evaluation of Speech Quality (PESQ) \cite{rix2001perceptual}, Scale-Invariant Signal-to-Distortion Ratio (SI-SDR)\cite{le2019sdr}, Short-Time Objective Intelligibility (STOI)\cite{taal2010short} and Noresqa-MOS (N-MOS) \cite{noresqamos}, which was a much better candidate than Noresqa \cite{manocha2021noresqa} for this kind of degradation. All the metrics have been computed on the test set for each benchmarked model. Speech enhancement being a one-to-many problem, these results should be analyzed cautiously. Indeed, a plausible signal with perfect intelligibility but still different from reference would be misjudged by the metrics. Note that these metrics are intrusive, since they require groundtruth audio. Generally speaking, speech quality assessment still lacks objective and non-intrusive evaluation metrics, although recent work such as  \cite{manocha2021noresqa} may be part of the solution. This observation is confirmed by \cite{vinay2022evaluating} which points out that current objective metrics are questionable.

\begin{table}[ht!]
    \centering
    \begin{adjustbox}{max width=0.5\textwidth,max totalheight=1.2\textheight}
    \begin{tabular}{|l||l|l|l|l|}
     \hline
      \diagbox[width=10em, height=0.6cm]{Speech}{Metrics}  &  PESQ  &  SI-SDR  &  STOI & N-MOS  \\
      \hline
      Simulated In-ear  &  \textbf{2.42 (0.34)}  &  8.4 (3.7)  &  0.83 (0.05) & 2.57 (0.58) \\
      Audio U-net \cite{kuleshov2017audio}   &  2.24 (0.49)  &  \textbf{11.9 (3.7)}  & 0.87 (0.04) & 2.59 (0.44) \\
      Hifi-GAN v3\cite{kong2020hifi}   &  1.32 (0.16)  & -25.1 (11.4)  & 0.78 (0.04) & 3.70 (0.68)\\
      Seanet  \cite{tagliasacchi2020seanet} &  1.92 (0.48)  &  11.1 (3.0)  &  \textbf{0.89 (0.04)} & \textbf{4.25 (0.28)} \\
      Streaming-Seanet  \cite{li2021real} &  2.01 (0.46) &  11.2 (3.6)  &  \textbf{0.89 (0.04)} & 3.91 (0.60) \\
      EBEN (ours)    &  2.08 (0.45)  & 10.9 (3.3)  & \textbf{0.89 (0.04)} & 4.02 (0.39) \\
      \hline
    \end{tabular}
    \end{adjustbox}
     \caption{PESQ/SI-SDR/STOI on test set. Significantly best values (acceptance=0.05) are in \textbf{bold}.}
	\label{tab:objective}
\end{table}

Even though purely reconstructive approaches have a clear advantage when evaluated on comparative metrics,  Kuleshov's model \cite{kuleshov2017audio} does not prevail on STOI, which is the comparative metric that is the most correlated with human evaluation for our specific task, as shown in \ref{sec:correlation}. Looking at these results, we could say that best performing models for STOI are either Seanet, Streaming-Seanet or EBEN.

\subsubsection{Frugality study}

Enhancing performances need to be qualified by the model's latency and heaviness to take deep learning from hype to real-world applications. Indeed, the bandwidth extension is applicable for a two-way communication device, if latency is roughly smaller than $20$~ms as claimed in \cite{lezzoum2016echo}. The total number of parameters influencing the memory space should also be reduced. Therefore, we reported on Tab.~\ref{tab:frugal} :
\begin{itemize}
 \item $P_{gen}$: The total number of parameters for the generator, including non-trainable parameters like PQMF-bank for EBEN. For other methods, preprocessing parameters like the mel windows are not counted.
 \item $P_{dis}$: The total number of parameters for discriminators.
 \item $\tau$: The latency corresponding to the generator's forward pass during inference (no gradients are calculated). We carefully synchronized GPU to account for any asynchronous execution and chose the fastest kernels by enabling the cudnn benchmark. The reported measures are averaged over 10000 points.
 \item $\delta$: The maximum memory allocation used during inference measured with \verb+torch.cuda.max_memory_allocated+.
\end{itemize}
$\delta$ and $\tau$ are given for a single one-second sample.

\begin{table}[ht!]
    \centering
    \begin{tabular}{|l||l|l|l|l|}
     \hline
      \diagbox[width=10em, height=0.6cm]{Speech}{Parameters}  & $P_{gen}$ & $P_{dis}$ & $\tau$ (ms)& $\delta$ (MB)\\
      \hline
      Audio U-net \cite{kuleshov2017audio}  & 71.0 M &  $\emptyset$ & 37.5 & 1117.3\\ 
      Hifi-GAN v3\cite{kong2020hifi} & 1.5 M &  70.7 M & \textbf{3.1} & 22.2\\  
      Seanet  \cite{tagliasacchi2020seanet} &  8.3 M &   56.6  M & 13.1 & 89.2 \\  
      Streaming-Seanet  \cite{li2021real} &  \textbf{0.7 M} &   56.6  M & 7.5 & \textbf{10.9} \\  
      EBEN (ours) &  1.9 M &  \textbf{27.8 M}& 4.3 & 20.0  \\  
      \hline
    \end{tabular}
     \caption{Parameters, latencies and memory usage of models}
	\label{tab:frugal}
\end{table}

Tab.~\ref{tab:frugal} nuances the simple study of model parameters. Indeed, neither $\tau$ nor $\delta$ linearly depends on the number of parameters. They are also influenced by models' depth, embedding width, and hyperparameters that will, for instance, determine the choice of the convolution algorithm (Winograd, FFT, GEMM). Thanks to the reduction operated by PQMF filtering, EBEN is the lightest proportionally to its parameters and one of the fastest networks. It is also more than 3 times faster to infer and 4 times lighter than Seanet  \cite{tagliasacchi2020seanet}.

\subsection{Subjective evaluation}
\label{sec:subjective}

\subsubsection{Visual inspection of spectrograms}

\begin{figure*}[ht!]
  \centering
  \centerline{\includegraphics[width=18.5cm]{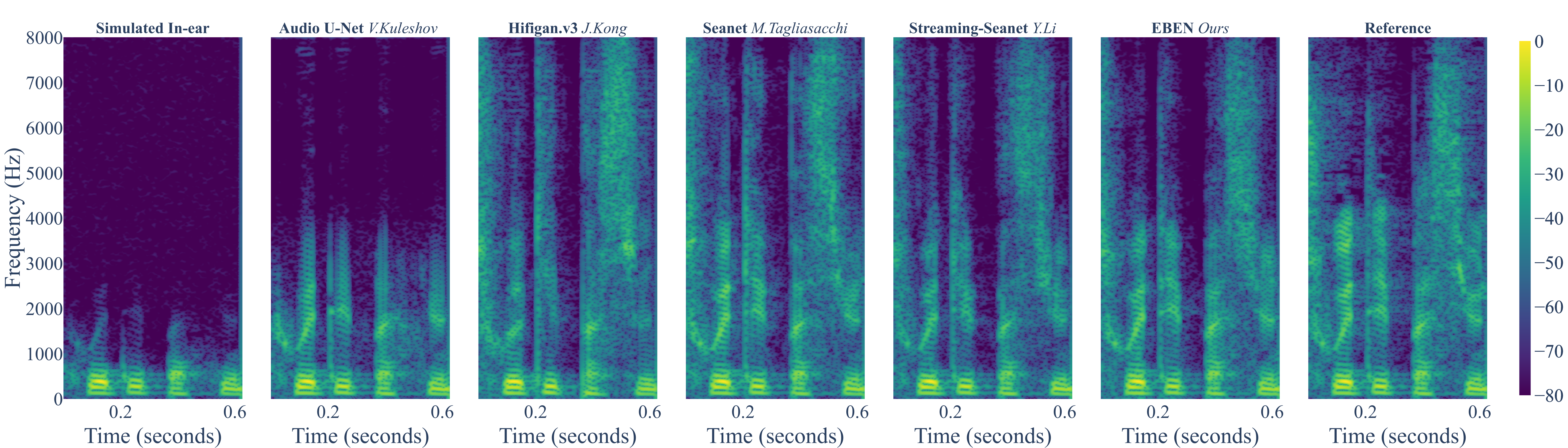}}
\caption{Spectrograms of various bandwidth extension models sandwiched by the simulated in-ear and the reference signals.}
\label{fig:spec}
\end{figure*}

To visually assess and compare the obtained results with each trained model, Fig.~\ref{fig:spec} shows some spectrograms obtained from the testing set. It can be observed that a purely reconstructive approach \cite{kuleshov2017audio} is not sufficient to produce high frequencies. Indeed, when low frequency information is insufficient, the model predicts the mean of speech signals, which is zero. Among generative approaches, our method is competitive. Indeed, EBEN reconstructs a fair amount of formants and minimizes artifacts. As a comparison, Hifi-GAN v3\cite{kong2020hifi} and Streaming-Seanet \cite{li2021real} are not as efficient for harmonic reconstruction. Seanet \cite{tagliasacchi2020seanet}seems to be the closest to the reference's spectrogram. All approaches were able to get rid of the additive Gaussian noise. Some additional zoomable spectrograms confirming these observations are available at \url{https://jhauret.github.io/eben/}.

\subsubsection{MUSHRA study}

We conducted a subjective comparative evaluation of the different trained models using the MUltiple Stimuli with Hidden Reference and Anchor \cite{series2014method} (MUSHRA) methodology. According to the MUSHRA specification, a rating scale ranging from 0 to 100 has been used; the higher, the better. A total of $56$ samples were rated by 170 participants, corresponding to 7 audios enhanced by five different networks, plus the hidden reference and a hidden low anchor (corresponding to an untrained EBEN network) and the simulated in-ear signal. Participants were recruited by e-mail to complete one of two available tests on the GoListen platform \cite{barry2021go} : MUSHRA-Q which allows to rank methods for produced sound quality, and MUSHRA-U, which aims at ranking methods for ease of speech understanding. Ease of understanding is linked with notions of phonetic confusion and intelligibility (but is not equivalent to standard listening test to assess intelligibility, such as the Modified Rhyme Test \cite{house1963psychoacoustic}), while audio quality reflects the naturalness and listening comfort. For both tests, we recorded participants' hearing status and type of sound reproduction system to retain 69 participants over 88 for MUSHRA-U and 66 over 82 for MUSHRA-Q. We also followed the two post-screening phases recommended by the International Telecommunication Union (ITU) \cite{series2014method} to retain only participants who provided consistent ratings:

\begin{itemize}
 \item \underline{Stage 1 post-screening:} \textit{A listener should be excluded from the aggregated responses if he or she rates the hidden reference condition for at least $15\%$ of the test items lower than a score of 90.}
 \item \underline{Stage 2 post-screening:} \textit{Exclude subjects whose individual grades fall outside 1.5 $\times$ the upper or lower bound of the IQR of the aggregated listeners for at least $25\%$ of the test items.\\}
\end{itemize}

\begin{figure}[htb]
  \centering
  \centerline{\includegraphics[width=0.9\linewidth]{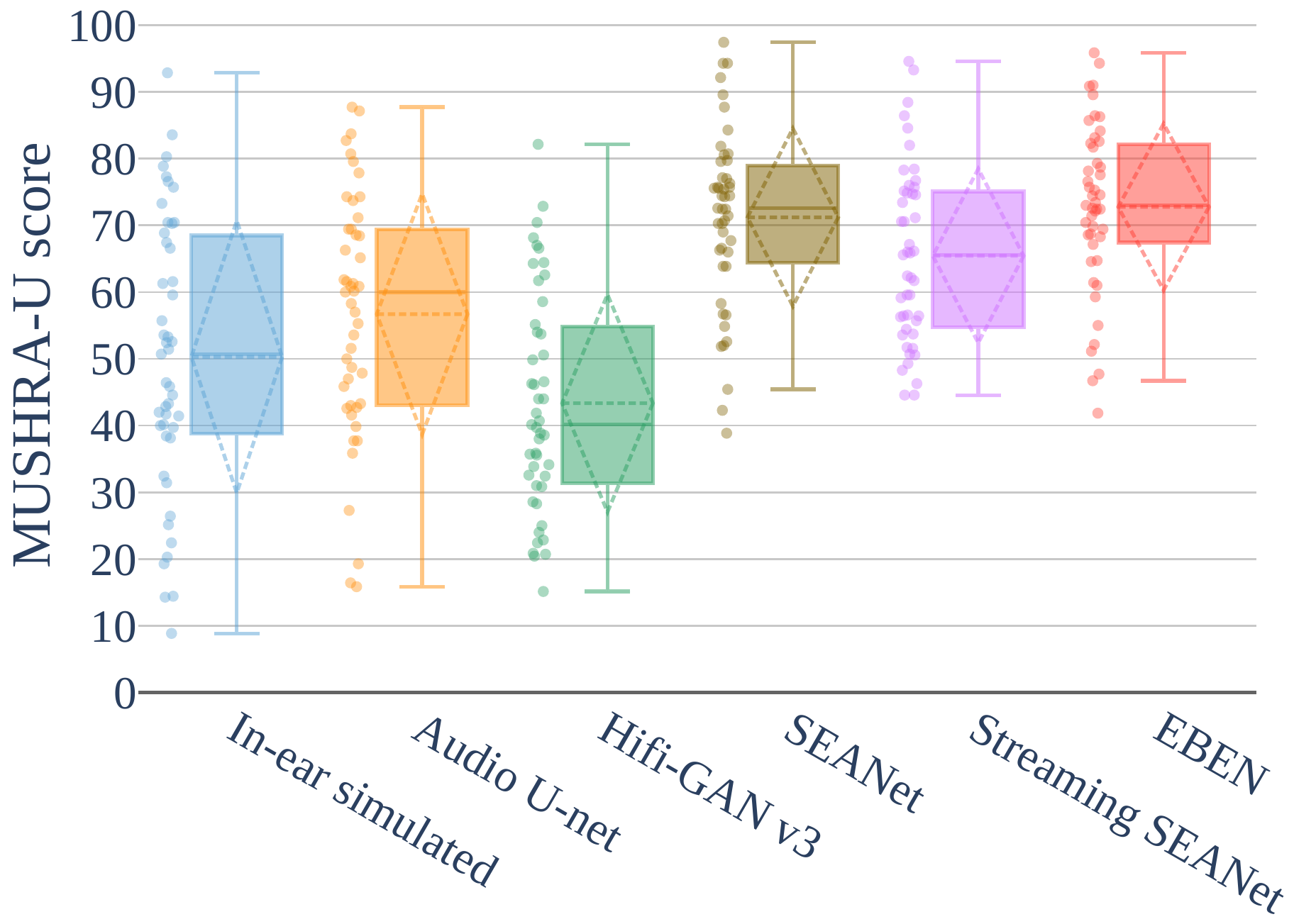}}
\caption{MUSHRA-U : statistical distributions of scores obtained with a MUSHRA procedure for the ranking of perceived ease of understanding across trained models.}
\label{fig:mushraI}
\end{figure}

\begin{figure}[htb]
  \centering
  \centerline{\includegraphics[width=0.9\linewidth]{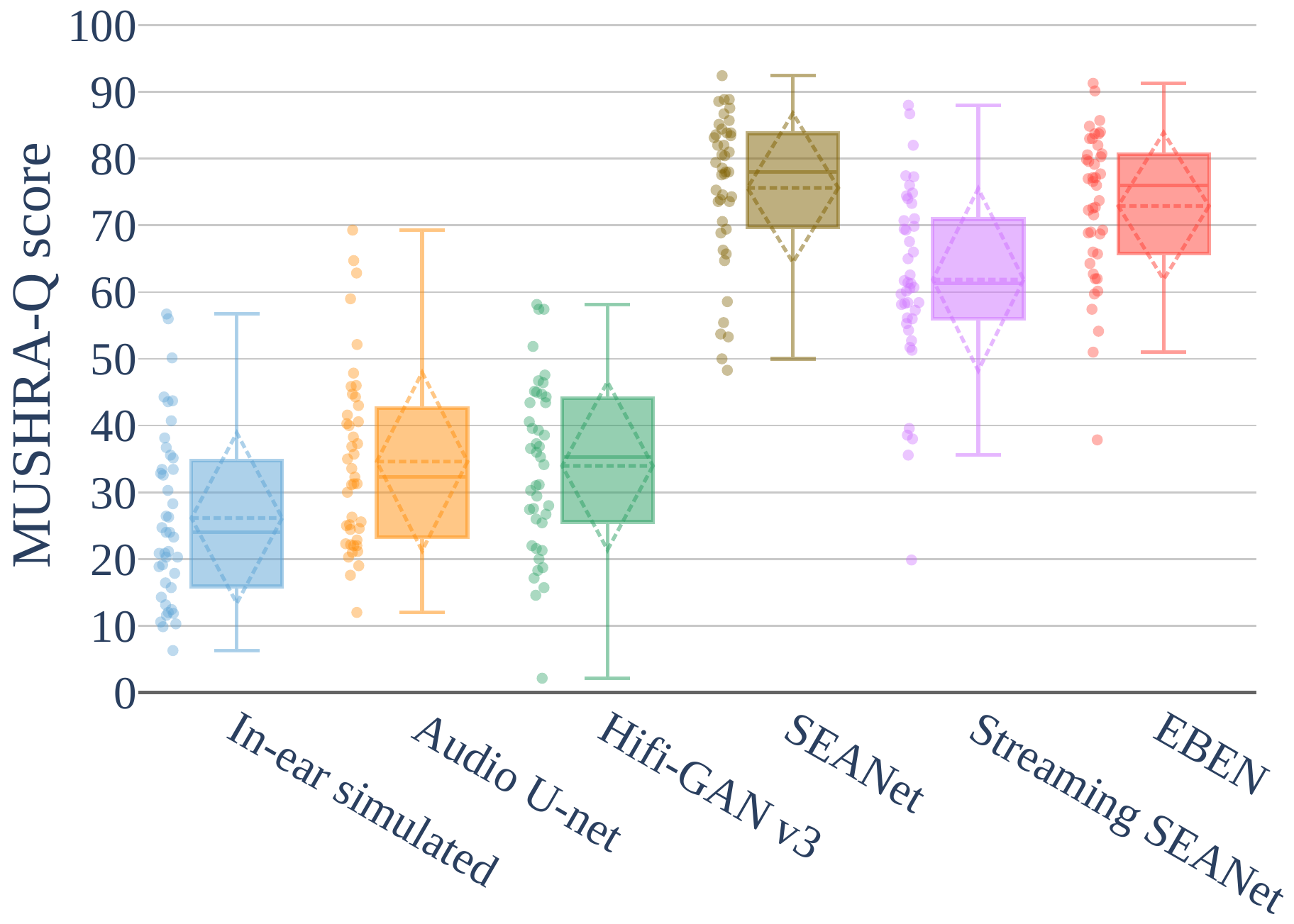}}
\caption{MUSHRA-Q : statistical distributions of scores obtained with a MUSHRA procedure for the ranking of perceived sound quality across trained models.}
\label{fig:mushraQ}
\end{figure}

After applying those two criteria, we retain 47/88 for MUSHRA-U and 43/82 for MUSHRA-Q. The overall age repartition is as follows: $37\%$ are below 27 years old, $24\%$ are above 50, and $39\%$ between 27 and 50 years old. We found no statistically significant differences in ratings between the age categories. The distribution of obtained gradings are shown Fig.~\ref{fig:mushraI} and Fig.~\ref{fig:mushraQ}.

The statistical distributions have been studied using a non-parametric Friedmann Analysis of Variance to confirm the statistical significance of the results. The obtained p-values are lower than $1e-20$, demonstrating that there are significant differences, both in terms of quality and intelligibility among tested approaches. This made it possible to perform a post-hoc Nemenyi-Friedmann analysis, in order to assess the 2-to-2 independence of the distributions. The obtained results show that the EBEN approach ranks first ex aequo with Seanet in terms of quality and ease of understanding (no statistically significant difference between EBEN and Seanet, p-value $>0.5$) and that these two methods significantly outperform the second best approach Streaming Seanet (p-value $<$ 0.005).

\section{Discussion}
\label{sec:discussion}

\subsection{PQMF insights}
\label{sec:pqmfinsights}
PQMF banks are helpful for a wide range of tasks, including audio equalization, noise reduction, or compression, e.g., by reducing the bit rate on sparser bands. This work has used the PQMF analysis outputs to speed up the inference by taking advantage of the decimation operator. The multiband representation has the same dimensionality as the original signal but is condensed along the time axis and extended along channels, allowing parallel computing. Also, by the very nature of our problem, some frequency bands of the input signal do not contain any information, and we can drop them. Furthermore, generating bands reduces redundancy, leading again to a reduction in computational complexity. Finally, it allows the design of discriminators for EBEN that act only where bandwidth extension is needed.

Along with EBEN source code, we also provide a modern and efficient implementation of the PQMF analysis and synthesis with native Pytorch functions, using only strided convolutions and strided transposed convolutions.

\subsection{EBEN's Configurability}
\label{sec:configurability}
We called our work " Configurable" because several aspects of EBEN's architecture can be adapted to address different BCM degradations. The corresponding hyperparameters are:

\begin{itemize}
 \item $M$: The number of PQMF bands. It has a direct impact on the width of each band's frequency bandwidth. Higher values of this parameter enable finer control over the bands, enhancing precision. Furthermore, it influences the downsampling factor. Raising the number of bands reduces the computational burden on the network, but it may result in some loss of temporal resolution. As pointed in \cite{zaiem2023fine}, downsampling proves to be the most effective option when finetuning models for faster inference, as it offers significant computational gains with minimal performance trade-offs. Moreover, when the number of bands is considerably large, the empirical equation $N=8M$ may require a convolution in the Fourier domain.

 \item $P$: The number of informative PQMF bands sent to the generator. Given the sampling frequency $F_s$ and the cutoff frequency of the low-pass filter $F_c$, the factor $\frac{P}{M}$ should be just above the reduced cutoff frequency of the degradation $\frac{2F_c}{F_s}$. In this way, all the information remaining in the low-pass filtered signal is captured and the high-frequency noise is discarded. We carried out several ablation studies with EBEN, by varying the number of $P$ bands for the specific degradation described in this article, using a constant number of bands $Q$ sent to the discriminators. By performing these ablation studies, we were able to determine that the inclusion of informationless bands significantly degrades the objective metrics obtained for the enhanced signals when some high-frequency noise is present, but has a minimal effect when the noise is absent. For a pure bandwidth extension task from $Fc = 4$~kHz to $F_c = 16$~kHz, when there is absolutely no content above the initial half sampling frequency, a value of $P = 1$ is also enough to obtain excellent results (STOI = 0.93 / N-MOS = 4.23 on test set after enhancement). Those conclusions of course heavily depend on the kind of degradation, which further highlights the benefits of a configurable approach.

 \item $Q$: The number of PQMF bands sent to the discriminators to be refined. In fact, the very first bands should contain clean low frequencies of the speech signal and would only require a slight modification by the MelGAN discriminator to suppress physiological noise. On the contrary, the last bands, which suffer from information loss, must be filled by the generator network, pushed to do so thanks to the PQMF discriminators. We also carried out several ablation studies with EBEN, by varying $Q$ in $[1;3]$ for a fixed value of $P = 1$. This study allows to draw a clear relationship between the degree of band refinement and the corresponding enhancement performance: as the number of refined bands decreases, the enhancement performance diminishes accordingly. We believe that discriminator configurability is once again beneficial for specific enhancement objectives in targeted frequency bands. Indeed, some BCMs exhibit a very sharp rolloff. In these cases, $P$ and $Q$ should be chosen to be complementary. On the other hand, other kinds of BCMs exhibit a smaller high-frequency rolloff. In such scenarios, EBEN's configurability allows $Q$ to be chosen so that the frequency bands input to the PQMF discriminators overlap with the $P$ frequency bands fed to the generator. This design allows to refine the upper bands that are degraded, even if there is still some residual speech content that can be useful to the generator for the bandwidth extension task.

\end{itemize}

\subsection{Correlation between subjective and objective metrics}
\label{sec:correlation}
\begin{figure}[htb]
  \centering
  \centerline{\includegraphics[width=0.9\linewidth]{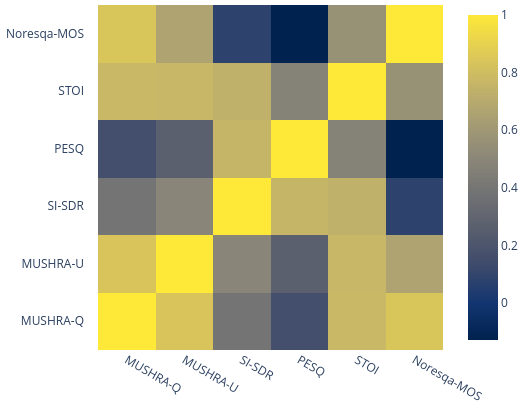}}
\caption{Pearson product-moment correlation coefficients of objective and subjective metrics}
\label{fig:corrcoef}
\end{figure}

General purpose intrusive (or comparative) metrics for assessing speech intelligibility and quality are far from perfect. However, the  existing literature still relies on those common set of metrics for evaluation purposes for lack of anything better. In the present study for example, the PESQ metric ranks the Audio U-net \cite{kuleshov2017audio} approach as the best method, which contradicts the results obtained with subjective evaluation. Having MUSHRA test results and the various metrics at our disposal enables us to perform a correlation analysis between the objective metrics and the subjective tests, in order to provide an overview of which metrics are best/less suited to the bandwidth extension task. The results are shown in Fig~.\ref{fig:corrcoef}.

We can deduce from the coefficients shown in Fig~.\ref{fig:corrcoef} that SI-SDR and PESQ are not suitable to evaluate the quality of bandwidth extension methods because of their poor correlation with MUSHRA. On the other hand, this analysis allows to conclude that STOI and Noresqa-MOS are two relevant indicators. Hence, the pseudo-ranking of the relevance level of the metrics for our specific use case is Noresqa-MOS$\approx$STOI$>>$SI-SDR$>$PESQ. It is also noteworthy that Noresqa-MOS is more correlated with MUSHRA-Q than with MUSHRA-U. This seems logical since Noresqa-MOS is built to predict quality.

\subsection{Accordance of the synthetic data generation}
\label{sec:phirandom}
In this part, we would like to reflect on the relevance of the generated synthetic data to tackle some real in-ear captured speech. To do this, we used two EBEN models with the same configuration. One was trained to reverse the $\Psi_{fixed}$ degradation and discussed in \ref{sec:experiments}, while the other was trained on $\Psi_{random}$ which is closer to the real degradation by design. We chose EBEN among the other approaches, but this section is independent to the model choice. Although we have selected two models that perform well with in-distribution data, the results obtained on Tab.~\ref{tab:robustness} show that neither model is able to significantly enhance the raw in-ear speech signal according to objective metrics.

\begin{table}[ht!]
\centering
\begin{tabular}{|p{31mm}||l|l|l|l|}
\hline
Speech& PESQ & SI-SDR & STOI & N-MOS \\ \hline
In-ear enhanced via EBEN trained on $\Psi_{fixed}$  & 1.16 & -37.4 & 0.51 & 3.82 \\ \hline
In-ear enhanced via EBEN trained on $\Psi_{random}$ & 1.28 & -41.6 & 0.53 & 3.80  \\ \hline
Raw In-ear  & 1.5  & -37.0   &  0.56 & 3.33           \\ \hline
\end{tabular}
\caption{EBEN's ability to enhance real data according to different training sets}
\label{tab:robustness}
\vspace{-0.1cm}
\end{table}

Efforts to get closer to $\Psi_{smoothed\_median}$ for $\Psi_{random}$ did not pay off because the complex degradation cannot be accurately simulated by a linear transfer function. Rather, the degradation is likely non-linear. Moreover, the additive physiological and frictional noise is time-dependent, making the assumption of a linear time-invariant system untrue in practice. Therefore, instead of investing a significant amount of time and effort to create a suitable simulation model, we will try to use real data in our future works. Indeed, the data-driven nature of deep learning suggests that the training set should be based on real data: we are in the process of building and releasing a complete BCM recording dataset.

\section{Conclusion}
\label{sec:conclusion}
We presented Configurable EBEN: a state-of-the-art, real-time compatible, and lightweight neural network architecture to address the problem of unimodal enhancement of speech signals captured with noise-resilient body-conduction microphones. The main challenge encountered with these unconventional microphones is the need to achieve a bandwidth extension of the raw captured signals. We therefore designed EBEN to be fully configurable for the bandwidth enlargement needed. We specifically proposed a multiband approach, where the enhancement is solely conditioned on the first $P$ informative bands, and the adversarial training is mainly targeted to enhance $Q$ bands over a total of $M$ bands through newly designed discriminators. Furthermore, this multiband decomposition -- which is using Pseudo Quadrature Mirror Filter bank -- enables a reduction of the feature dimensionality from the very first layer of the encoder. This benefits streaming compatibility, because fewer computations are required during the forward pass and reduce data redundancy. Those findings are supported by extensive experimentation and comparisons with existing models. These experiments demonstrate that EBEN is competitive in many aspects, including enhancement performance, latency, and memory footprint. EBEN is therefore a good compromise between frugal AI requirements and speech enhancement performance; ready to be trained on a real BCMs dataset.

\textbf{Acknowledgements}: This work has been partially funded by the French National Research Agency under the ANR Grant No. ANR-20-THIA-0002. This work was also granted access to the HPC/AI resources of [CINES / IDRIS / TGCC] under the allocation 2022-AD011013469 made by GENCI.

\bibliographystyle{IEEEtran}
\bibliography{eben.bib}

\begin{IEEEbiography}[{\includegraphics[width=1in,height=1.25in,clip,keepaspectratio]{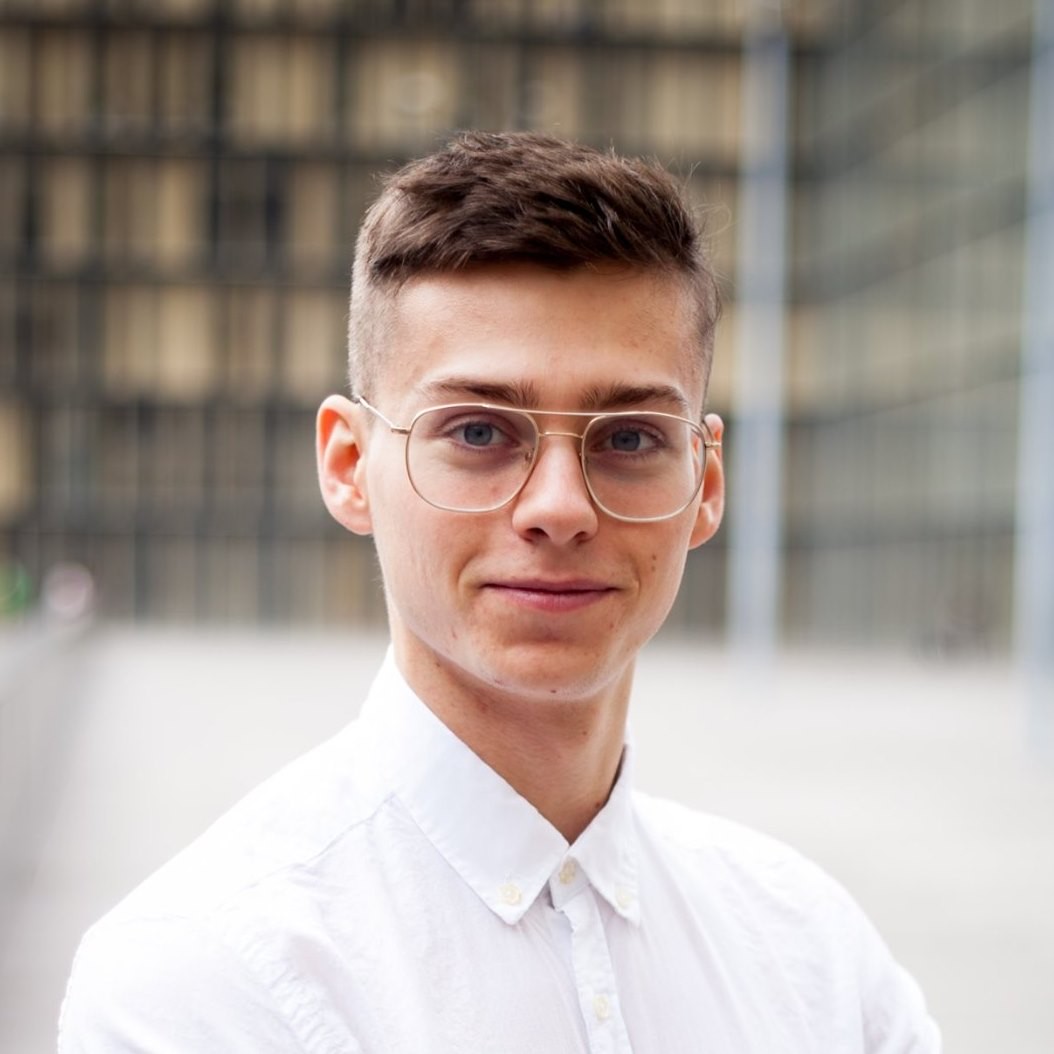}}]{Julien Hauret}
\input{bios/Julien_Hauret.tex}\end{IEEEbiography}

\begin{IEEEbiography}[{\includegraphics[width=1in,height=1.25in,clip,keepaspectratio]{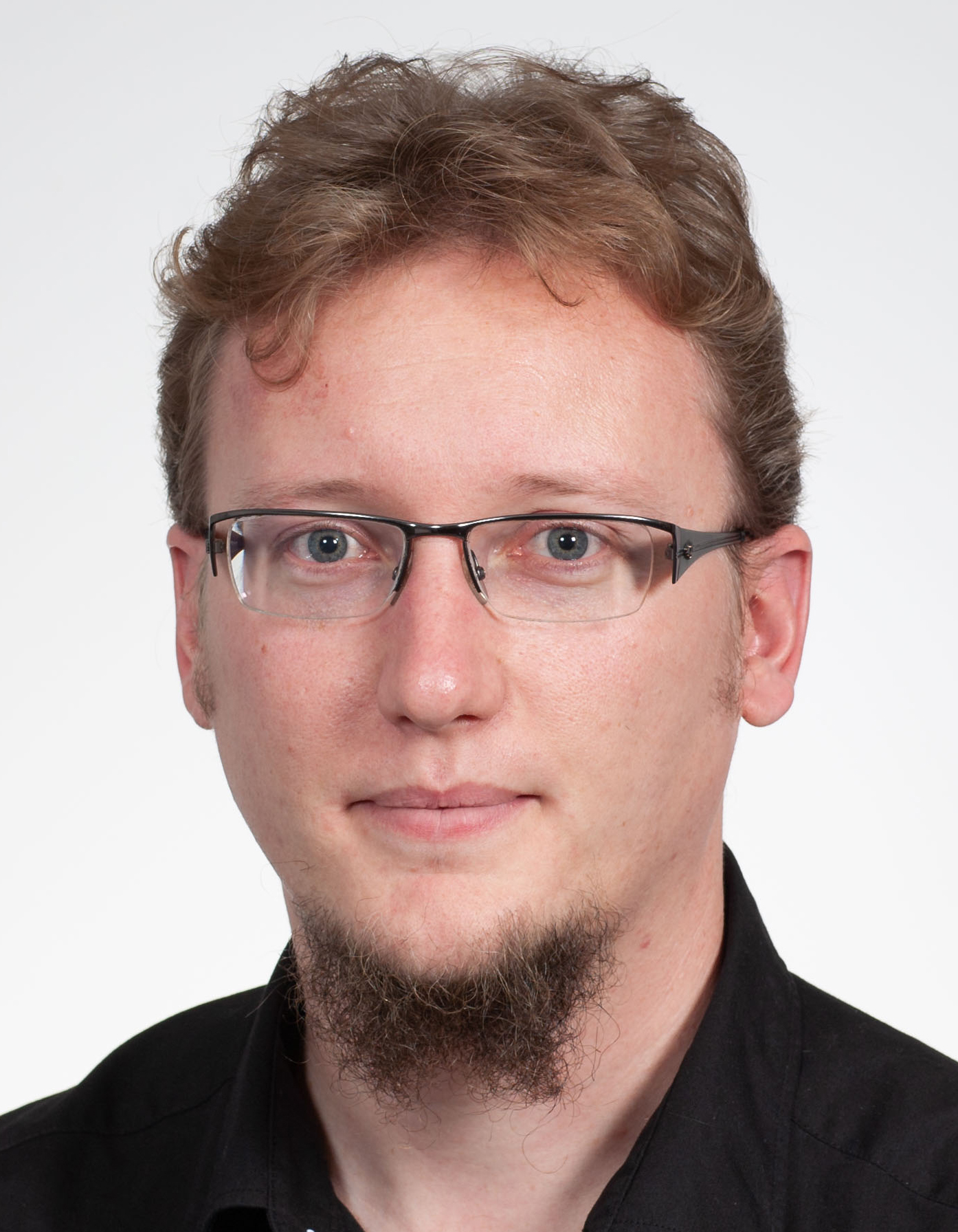}}]{Thomas Joubaud}
\input{bios/Thomas_Joubaud.tex}\end{IEEEbiography}

\begin{IEEEbiography}[{\includegraphics[width=1in,height=1.25in,clip,keepaspectratio]{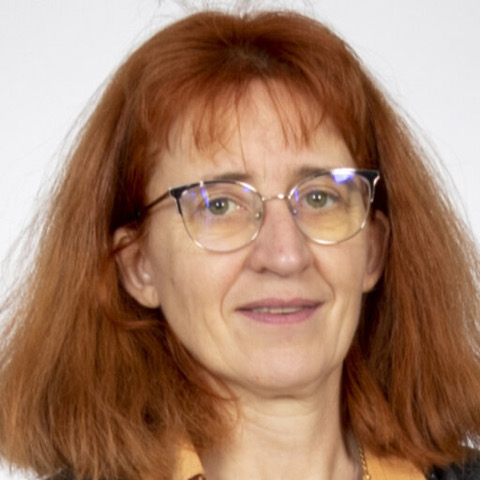}}]{Véronique Zimpfer}
\input{bios/Veronique_Zimpfer.tex}\end{IEEEbiography}

\begin{IEEEbiography}[{\includegraphics[width=1in,height=1.25in,clip,keepaspectratio]{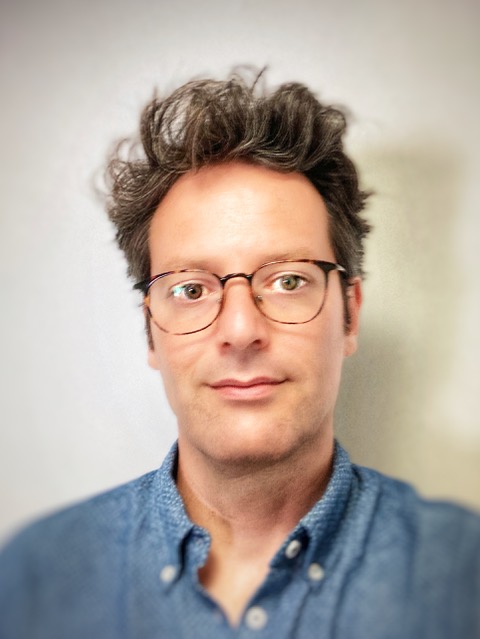}}]{Éric Bavu}
\input{bios/Eric_Bavu.tex}\end{IEEEbiography}

\end{document}

%% file: bios/Julien_Hauret.tex
is a PhD candidate at Cnam Paris, pursuing research in machine learning applied to acoustics. He holds two MSc from ENS Paris Saclay, one in Electrical Engineering (2020) and the other in Applied Mathematics (2021). He has a strong research training as evidenced by his internships at Columbia University and the French Ministry of the Armed Forces. He also lectures on algorithms and data structures at the École des Ponts ParisTech. His research focuses on the use of deep learning for speech enhancement applied to body-conducted speech. With a passion for interdisciplinary collaboration, Julien aims to improve human communication through technology.

%% file: bios/Thomas_Joubaud.tex
 is a Research Associate at the Acoustics and Soldier Protection department within the French-German Research Institute of Saint-Louis (ISL), France, since 2019. In 2013, he received the graduate degree from Ecole Centrale Marseille, France, as well as the master's degree in Mechanics, Physics and Engineering, specialized in Acoustical Research, of the Aix-Marseille University, France. He earned the Ph.D. degree in Mechanics, specialized in Acoustics, of the Conservatoire National des Arts et Métiers (Cnam), Paris, France, in 2017. The thesis was carried out in collaboration with and within the ISL. From 2017 to 2019, he worked as a post-doctorate research engineer with Orange SA company in Cesson-Sévigné, France. His research interests include audio signal processing, hearing protection, psychoacoustics, especially speech intelligibility and sound localization, and high-level continuous and impulse noise measurement.

%% file: bios/Veronique_Zimpfer.tex
 is a Scientific Researcher at the Acoustics and Soldier Protection department within the French-German Research Institute of Saint-Louis (ISL), Saint-Louis, France, since 1997. She holds a M.Sc in Signal Processing from the Grenoble INP, France and obtained a PhD in Acoustics from INSA Lyon, France, in 2000. Her expertise lies at the intersection of communication in noisy environments and auditory protection. Her research focuses on improving adaptive auditory protectors, refining radio communication strategies through unconventional microphone methods, and enhancing auditory perception while utilizing protective gear.

%% file: bios/Eric_Bavu.tex
is a Full Professor of Acoustics and Signal Processing at the Laboratoire de Mécanique des Structures et des Systèmes Couplés (LMSSC) within the Conservatoire National des Arts et Métiers (Cnam), Paris, France. He completed his undergraduate studies at École Normale Supérieure de Cachan, France, from 2001 to 2005. In 2005, he earned an M.Sc in Acoustics, Signal Processing, and Computer Science Applied to Music from Université Pierre et Marie Curie Sorbonne University (UPMC), followed by a Ph.D. in Acoustics jointly awarded by Université de Sherbrooke, Canada, and UPMC, France, in 2008. He also conducted post-doctoral research on biological soft tissues imaging at the Langevin Institute at École Supérieure de Physique et Chimie ParisTech (ESPCI), France. Since 2009, he has supervised six Ph.D. students at LMSSC, focusing on time domain audio signal processing for inverse problems, 3D audio, and deep learning for audio. His current research interests encompass deep learning methods applied to inverse problems in acoustics, moving sound source localization and tracking, speech enhancement, and speech recognition.